\documentclass[manuscript,screen]{acmart}

\AtBeginDocument{%
  }

\usepackage{url}
\usepackage{framed}

\newtheorem{definition}{Def.}

\begin{document}

\title{Hot Fixing Software: A Comprehensive Review of Terminology, Techniques, and Applications}

\author{Carol Hanna}
\email{carol.hanna.21@ucl.ac.uk}
\orcid{0009-0009-7386-1622}
\affiliation{%
  \institution{University College London}
  \city{London}
  \country{United Kingdom}
}

\author{David Clark}
\email{david.clark@ucl.ac.uk}
\orcid{0000-0002-7004-934X}
\affiliation{%
  \institution{University College London}
  \city{London}
  \country{United Kingdom}
}

\author{Federica Sarro}
\email{f.sarro@ucl.ac.uk}
\orcid{0000-0002-9146-442X}
\affiliation{%
  \institution{University College London}
  \city{London}
  \country{United Kingdom}
}

\author{Justyna Petke}
\email{j.petke@ucl.ac.uk}
\orcid{0000-0002-7833-6044}
\affiliation{%
  \institution{University College London}
  \city{London}
  \country{United Kingdom}
}


\begin{abstract}
A hot fix is an unplanned improvement to a specific time-critical issue deployed to a software system in production. 
While hot fixing is an essential and common activity in software maintenance, it has never been surveyed as a research activity.
Thus, such a review is long overdue. 
In this paper, we conduct a comprehensive literature review of work on hot fixing. 
We highlight the fields where this topic has been addressed, inconsistencies we identified in the terminology, gaps in the literature, and directions for future work.
Our search concluded with 91 articles on the topic between the years 2000 and 2022.
The articles found encompass many different research areas such as log analysis, runtime patching (also known as hot patching), and automated repair, as well as various application domains such as security, mobile, and video games.
We find that many directions can take hot fix research forward such as unifying existing terminology, establishing a benchmark set of hot fixes, researching costs and frequency of hot fixes, and researching the possibility of end-to-end automation of detection, mitigation, and deployment. We discuss these avenues in detail to inspire the community to systematize hot fixing as a software engineering activity.
\end{abstract}




\maketitle

\section{Introduction}





Software maintenance is an essential activity in the software engineering life cycle~\cite{Pamunuwa2023}.
After a software system is deployed, a lot of engineering effort is directed to maintain it.
The maintenance activities ensure that both the functional and non-functional requirements of the system are upheld.
These activities are especially important because ensuring that a system is correct under all possible conditions before it is deployed, is generally not feasible in practice.
This is due to software testing being necessarily incomplete~\cite{Dijkstra1969} as well as the tight release deadlines in modern enterprises~\cite{tightRelease,testingChallengesFBISSTA2019,STchallengesICST23}.

Not all of the issues in production will have the same priority.
While most maintenance activities involve system upgrades and patches to improve the underlying software, some of the issues are critical and as such they require quick remediation.
These critical issues are often ones that cause severe degradation in performance, security vulnerabilities, or serious functional defects visible to the end user.
Patching these critical issues is often referred to as ``hot fixing''.

A traditional good fix for a software issue is expected to be correct, not break existing functionality, and not increase the technical debt of the system.
However, in the case of hot fixes these criteria do not necessarily apply.
Hot fixes need to remediate the unwanted symptoms of the critical issue as soon as possible~\cite{Truelove2021}.
In this case, a quick temporary fix is favored over a slower permanent fix that preserves the criteria~\cite{Chen2018}.
There is less emphasis on correctness under all conditions and more on the time it takes to generate a plausible patch that hides the critical symptom without breaking the system.
For these reasons, hot fixes for critical issues differ drastically from general patches.

The time criticality and temporary nature of hot fixing make the process of detecting when such fixes need to be applied, generating them, and deploying them not as systematic as other activities within software engineering.
These are the issues that stakeholders are only concerned with at the time when it is urgently required.
This is manifested in industrial practices and research activities on hot fixing, where the work is less established and the terminology is often inconsistent.

In this paper, we aim to aid in understanding the existing collective knowledge on hot fixing and in driving research in the area forward.
We have used a rigorous search to conduct the first-ever comprehensive literature review on the topic. 
Such a review is long overdue.
This paper makes the following contributions:
\begin{enumerate}
    \item A unified definition for \textit{hot fix} to align the terminology used in research on the topic, while keeping consistent with the existing body of work (Definition~\ref{def:hotfix}).
    \item A comprehensive survey of the literature on hot fixing that encapsulates terminology (Section~\ref{terminologyEvolution}), publication trends (Section~\ref{pubTrends}), techniques and applications (Section~\ref{techniquesAndApplications}).
    \item A detailed research agenda with the current open challenges in the area of hot fixing (Section~\ref{discussion}).
\end{enumerate}

We hope that these contributions will help drive research in the area of hot fixing forward.
Our website with a list of the included publications is available at \url{https://carolhanna01.github.io/hotfixes.github.io/}.\footnote{
All our raw data will be made available upon acceptance. 
}


\section{Terminology}
\label{terminologyEvolution}
The term \textit{hot fix} has had different usages in the literature~\cite{Hanna2023}.
Therefore, we first provide a brief history of the evolution of the term \textit{hot fix} and associated definitions found in previous work~\cite{Hanna2023}, before proposing a unified definition to streamline future research in the area.

Definitions in the research literature on hot fixing can be divided into two main categories.
The first regards a hot fix as a fix that needs to be deployed into a system in production during runtime without having to restart or reboot the system~\cite{Araujo2020}~\cite{Zhou2020}.
The second definition focuses on the criticality element of a hot fix.
In this regard, a hot fix is a time-critical fix that is temporary, small in size, and targets a specific issue in a system in production~\cite{Gupta2008}~\cite{Agarwal2014}.
Some domain-specific definitions exist as well in the fields of operating systems and information retrieval~\cite{Han2023}~\cite{Oosterhuis2021}.

Which of these definitions came first remains unclear.
However, the literature seems to suggest that the word \textit{hot} in the term signified the liveness of the system into which the patch was being deployed.
Thus, it is most likely that hot fixing was meant to describe the dynamic deployment of fixes into live systems.
From there, developers responsible for hot fixing activities began to evolve the term's meaning in other directions.
As a developer, when a fix needs to be deployed during run-time it must be an important and time-critical change otherwise it would simply be deployed within the planned software release cycle.
This interpretation with respect to timing has created two distinct definitions of the term, separating the literature into two fields.
While there can be an overlap between them as hot patches can sometimes be hot fixes and vice versa, the cross-pollination between the two communities remains limited.

An important point to notice is that most commonly the term \textit{hot fix} was associated with the time-criticality definition and \textit{hot patch} with the run-time definition.
We wish to follow the existing body of work and thus propose that the community use the term \textit{hot fix} for the time-criticality definition and \textit{hot patch} for the run-time definition moving forward.

The concept of time-criticality is nuanced and depends on the context.
Some work considers the time-criticality of an issue to be indicated by its severity, priority, and even how often it is reopened~\cite{criticalBugs}.
This might be subjective in some instances.
A time-critical issue might not be a breaking change necessarily but an issue that affects a specific important customer of the enterprise for example.
Thus, we leave the time-criticality part of our definition open to be \textit{templated}, based on the business needs.

Hot fixing as a software engineering activity usually falls outside the traditional software engineering life cycle.
As previously explained, this is due to the time constraint which results in quick development of workarounds, skipping extensive testing, and having deployment not wait until the next scheduled release.
From here, we can see that the criticality of the hot fix usually results in some form of exceptionality in its development process.

Hot patch refers to the runtime software patching activity, and is defined as follows by Islam et al.~\cite{Islam2023}: A \textit{hot patch}, also referred to as a \textit{runtime software patch}, ``aims to update a given software system while preserving running processes and sessions"~\cite{Islam2023}.

From here, one can consider a hot fix to be a \textit{phenomenon} in software engineering. This phenomenon usually breaks the traditional software engineering life cycle to address emergency issues in the system.
As for hot patching, this is more of a software \textit{technique} or a set of techniques that relate to runtime patching. 
Thus, the two terms \textit{hot patch} and \textit{hot fix} cannot be directly compared and should also not be used interchangeably. 

To streamline the literature on the topic, we propose to proceed with the following unified definition for the term \textit{hot fix}:\looseness=-1


\begin{framed}
\begin{definition}
\label{def:hotfix}
\noindent A \textbf{hot fix} is an unplanned improvement to a specific time-critical issue deployed to a software system 

in production.
\end{definition}
\end{framed}


\section{Survey Scope}
\label{scope}
To the best of our knowledge our survey is the first on hot fixing for software.
Islam et al. recently conducted a comprehensive study on runtime software patching which addresses current gaps in the literature and insights on future directions~\cite{Islam2023}.
Their survey details the state-of-the-art in runtime software patching with a scope that focuses on the deployment phase only which is the application of the patch itself into the end system.
More specifically, they look at dynamic/runtime patching techniques only for this step.
They do not look at critical patches specifically and instead look more generally at all runtime patching.
Since Islam et al.'s survey~\cite{Islam2023} covers previous work relating to the \textit{hot patch} category of definitions that we found in the literature (refer to Section~\ref{terminologyEvolution}) we exclude this line of work from the scope of our survey, and
focus its scope to hot fixing only per the definition that we proposed (Definition~\ref{def:hotfix}).

It is important to notice that our scope can still include some articles that use runtime patching, but only if the runtime patching technique specifically addresses \textbf{critical software issues}.
In that case, since the criticality property is met, the software activity would still be considered hot fixing and the technique used for it would be considered runtime patching. As such, articles of this kind are within our scope.
Moreover, studies on monitoring the fix post-deployment are excluded from our scope. This is because monitoring for post-deployment issues is not specifically related to hot fixing. For monitoring, the context in which code was integrated into production is less relevant.

We summarize the scope of our survey as follows:

\begin{framed}
\noindent\textbf{Scope:} Previous work is in the scope of this survey if it
\begin{enumerate}
  \item investigates the \textbf{detection} of critical software issues that hot fixes target \textit{at code level}; OR
  \item investigates the \textbf{repair} of critical software issues through the generation of hot fixes \textit{at code level}; OR
  \item investigates the \textbf{deployment} of hot fixes \textit{at code level}; OR
  \item empirically analyzes hot fixes in software systems.
\end{enumerate}
\end{framed}

\section{Survey Methodology}

\label{methodology}
To give a comprehensive analysis of the existing literature on hot fixing, we conducted a comprehensive literature search.
We set a rigorous methodology to ensure a literature review that would result in a comprehensive identification and analysis of existing work.
In this section, we detail such a methodology as well as the results that it derived.

\begin{table}
  \caption{Results of primary search for papers on hot fixes.} 
  \centering
  \label{tab:primaryResults}
  \begin{tabular}{lllrr}
    \toprule
    Keyword & 
    \multicolumn{4}{c}{`hot fix'}\\
    \midrule
    Source & Date of & Filters &Papers & Relevant\\
    & search && found & papers\\
    \midrule
    IEEE Xplore & 8/2/2023 & Full text \&  & 155 & 25\\
    && Metadata & & \\
    ACM Digital & 21/2/2023 & Title OR & 184 & 52\\
    Library && Abstract\\
    DBLP & 6/2/2023 & Default & 25 & 1\\
    Computer\\
    Science\\
    Bibliography \\
    ScienceDirect & 6/2/2023 & Default & 183 & 9\\
     & & abstract or \\
     & & author-\\
     & & specified\\
     & &  keywords\\
    \midrule
    \multicolumn{3}{l}{Total number of papers} & 547 & 87\\
    \midrule
    \multicolumn{4}{l}{Distinct number of papers} & 82\\
  \bottomrule
\end{tabular}
  \begin{tabular}{lllrr}
    Keyword & 
    \multicolumn{4}{c}{`hot patch'}\\
    \midrule
    Source & Date of & Filters & Papers & Relevant\\
    & search && found & papers\\
    \midrule
    IEEE Xplore & 9/2/2023 & Full text \&  & 116 & 25\\
    && Metadata & & \\
    ACM Digital & 17/2/2023 & Title OR & 98 & 24\\
    Library && Abstract\\
    DBLP & 9/2/2023 & Default & 21 & 11\\
    Computer\\
    Science\\
    Bibliography \\
    ScienceDirect & 9/2/2023 & Title & 240 & 3\\
     & & abstract or \\
     & & author-\\
     & & specified\\
     & &  keywords\\
    \midrule
    \multicolumn{3}{l}{Total number of papers} & 475 & 63\\
    \multicolumn{4}{l}{Distinct number of papers} & 54\\
  \midrule
  \midrule
  \multicolumn{4}{l}{\textbf{Total number of distinct papers on hot fixing}} & \textbf{136} \\
  \multicolumn{4}{l}{\textbf{Total in scope papers on hot fixing}} & \textbf{46} \\
  \bottomrule
\end{tabular}
\end{table}

\begin{table}
  \caption{Results of snowballing search for papers on hot fixes.} 
  \centering
  \label{tab:snowballResults}
  \begin{tabular}{lrr}
    \toprule 
    Search step & Total papers found & Relevant papers\\
    \midrule
     Round 1 & 985 & 23 \\ 
     \midrule
     Round 2 & 876 & 14 \\
     \midrule
     Round 3 & 396 & 3 \\
     \midrule
     Round 4 & 76 & 1 \\
     \midrule
     Round 5 & 26 & 0 \\
     \midrule
     \multicolumn{2}{l}{\textbf{Total number of new papers found}} & \textbf{41} \\
    \bottomrule
  \end{tabular}
\end{table}

\begin{table}
  \caption{Summary of results of literature review on hot fixes.} 
  \centering
  \label{tab:summaryResults}
  \begin{tabular}{lrr}
    \toprule 
    Search step & Total papers found \\
    \midrule
    Primary & 46 \\ 
    Snowballing & 41 \\ 
    Suggestions from authors & 4 \\ 
    \midrule
     \textbf{Papers on hot fixes for software} & 91 \\ 
  \bottomrule
\end{tabular}
\end{table}

Table~\ref{tab:primaryResults} presents the results of our primary search on keywords.
We started off with the keyword ``hot fix'' and subsequently also searched the keyword ``hot patch'' as we found that sometimes the words \textit{fix} and \textit{patch} are used interchangeably in the literature.
Variations of these two keywords including plurals and suffixes were included as well. 
We conducted our search using four computer science digital libraries: IEEE Xplore~\cite{ieee}, ACM Digital Library~\cite{acm}, DBLP Computer Science Bibliography~\cite{dblp}, and ScienceDirect~\cite{sciencedirect}.
For each of the search engines, the table presents the date on which the search was conducted, the filters used in the search, the total number of results given these filters, and finally the number of papers that we deemed as relevant.
The search was not restricted to a specific time-frame.
We looked at all papers published until the search dates presented in the paper.
The earliest publication we could find was from the year 2000 and the latest was 2022.
Thus, the time frame of this survey is 2000-2022.

At this stage in the review, we consider a paper to be relevant if it is in the domain of computer science and advances the knowledge on hot fixing software.
Using this criterion, we found 82 distinct papers using the keyword ``hot fix'' and 54 using the keyword ``hot patch''.
The primary search at this stage concluded in a total of 136 relevant distinct papers. 

Given these 136 relevant papers, we assessed their content more closely to better understand the scope that they cover. 
We found that 14 of these papers strictly address hot fixing within the context of project management.
Specifically, we found that these papers provide high-level tips for handling instances where hot fixing is required from the perspective of managing the full project.
In 19 of the papers, a definition for ``hot fix'' or ``hot patch'' is provided but the focus of the paper is not on this topic and thus there is no added novelty in this regard.
An additional 8 papers present scenarios for hot fixing a system.
However, the scenarios in these papers are provided as mere examples and are not the main subject of the paper.
For 25 of these papers, the term is used to refer to the runtime patching of bugs in software while not fulfilling the time-criticality criterium.
As this point, it became apparent that there is a clear inconsistency with the terminology.
Since runtime patching is not the topic of our survey these 25 papers were excluded as well.
Finally, we also removed 24 papers that just motivate the need for hot fixes in the software development life cycle without additional knowledge that would make them core papers on the topic.

At this stage of the literature review, we were left with 46 papers from the primary search which we deem as core papers on the topic.
The scope of our study was naturally born at this point which we will present in Section~\ref{scope}.

Following the primary search, we manually examined the bibliographies of all of the 46 papers that we deemed to fit the scope that we defined.
This process of snowballing on all of the papers referenced from each of the papers within the scope that we set was repeated until no more new papers were found.
The results of this snowball search are presented in Table~\ref{tab:snowballResults}. 
For each round of snowballing, we present the total number of papers found as well as the number of new distinct papers that fit within the scope of our survey.
Our snowballing process thus concluded after 5 rounds.


We present the final results of our literature review in Table~\ref{tab:summaryResults}.
We found 46 papers in the scope through the primary search and 41 additional papers in the snowballing search for a total of 87 papers.
After emailing the authors of these 87 papers to ask for feedback on the first draft of the survey, we received 4 additional relevant suggestions for papers to be included in the survey.
Overall, we ended up with 91 core papers which we describe in this survey.

After the paper collection process, we conducted a thematic analysis~\cite{Braun2012} as used in qualitative analysis research to categorize the papers found and effectively organize this paper.
We followed the six stages of thematic analysis which involved sifting through the papers, finding patterns among the collection, and naming the different categories.

The themes found correspond with the sections that we outline to present the existing work.
We initially divided the papers into two themes: empirical work and papers that present tooling.
From there, we further refined the type of empirical work depending on the data used: human studies, closed-source data, and open-source data.
As for the papers that cover tools, we refine them based on the type of tool.
We explain this in Section~\ref{techniquesAndApplications} and present the refinement in Figure~\ref{fig:existingWorkTaxonomy}.


\section{Publication Trends}
\label{pubTrends}
\begin{figure}
\center
\includegraphics[width=0.7\linewidth]{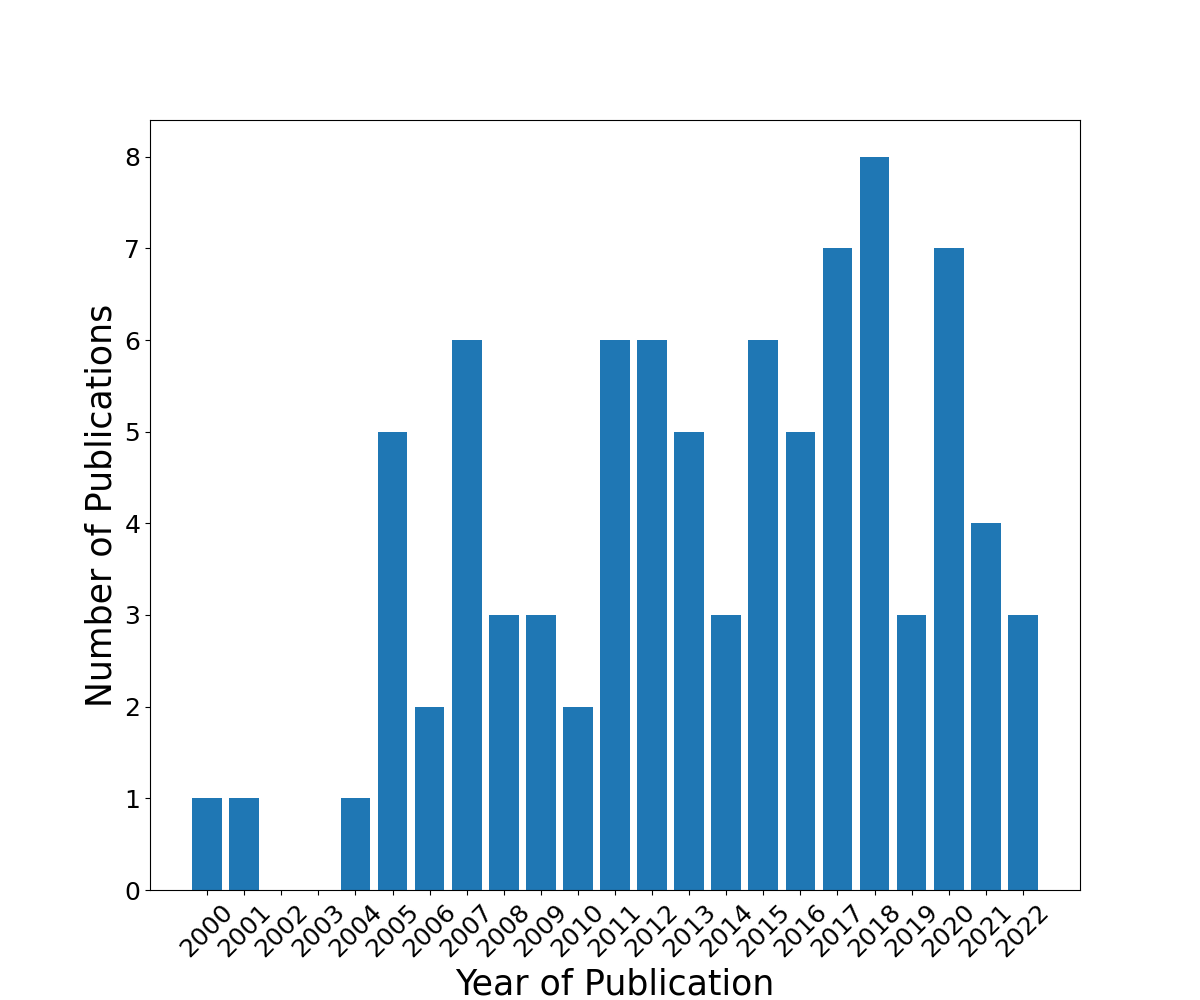}
  \caption{Publications on hot fixing software (2000-2022) over the years.}
  \label{fig:pubTrends}
\end{figure}

The trend of publications on the topic of hot fixing software is plotted in Figure~\ref{fig:pubTrends}.
The earliest relevant article on this topic that we were able to find was published in the year 2000.
We observe that publications on this topic started gain popularity after 2005. In more recent years, there has been a constant output of studies.
We hypothesize that this is due to the growing complexity of software systems which makes fixing urgent issues in production a difficult task.
As such, we can see that more research efforts have slowly started to be directed towards this software engineering activity.
Despite this, we observe that the number of annual publications remains very low with a maximum of 8 annual publications in 2018.
We hope that by reviewing the literature and streamlining what has already been done in tackling this topic, we can inspire future research in the area.

\begin{figure}
\includegraphics[width=0.6\linewidth]{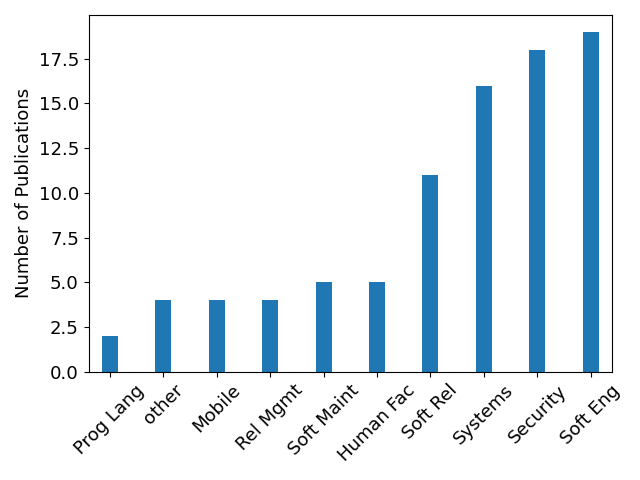}
  \caption{Research areas the surveyed publications pertain to (Programming Languages, Mobile, Release Management, Software Maintenance, Human Factors, Software Reliability, Systems, Security, Software Engineering, and Other).}
  \label{fig:researchAreas}
\end{figure}

We then assess the types of venues that the papers found were published in.
Since research on the topic spans multiple different fields as previously explained, we found a large variety of publication venues among the papers.
We plot the most popular research areas that the papers pertain to in Figure~\ref{fig:researchAreas}.
We found that the most popular were software engineering and security.


\begin{figure}
\centering
\includegraphics[width=0.5\linewidth]{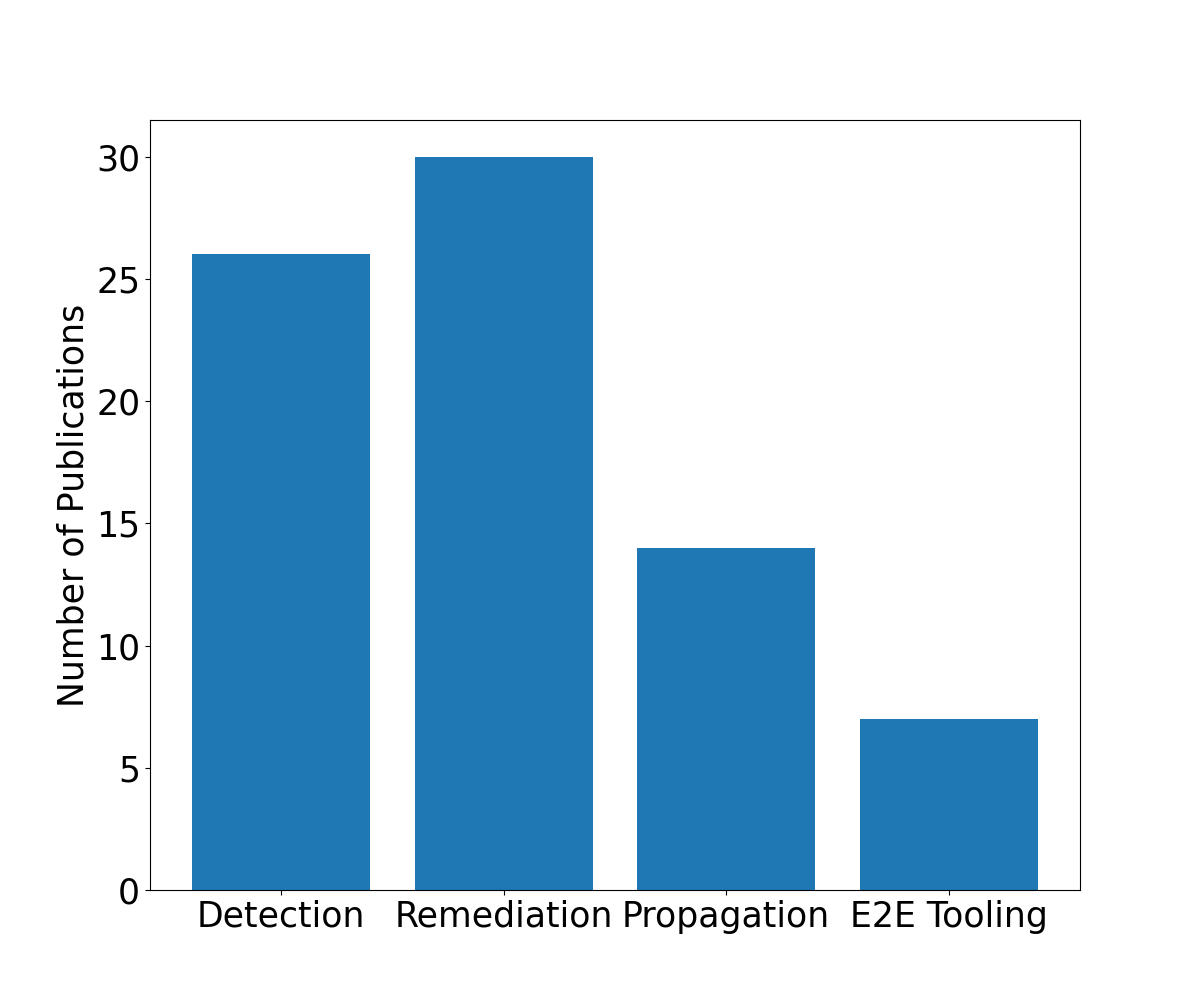}
  \caption{Number of publications for the different tooling types for hot fixing software. This includes detecting critical issues, remediation techniques for these issues, deployment strategies for their hot fixes into the target system, as well as end-to-end tools that encapsulate all three aforementioned stages.}
  \label{fig:toolingTypes}
\end{figure}

Following our thematic analysis, we were able to tag the papers into different categories.
Within the papers that include tooling, we analyzed the trend of tooling types and present this in Figure~\ref{fig:toolingTypes}.
We found that the most popular were critical issue remediation tools, closely followed by tools for detecting these critical issues, and finally around half that number for tooling to deploy the hot fixes for these critical issues into the target system.
We were only able to find a very limited number of end-to-end tooling that includes all three stages.
From the tools that generate patches, there wasn't a very big gap between those that can be deployed at runtime (45.8\%) and those that cannot (54.2\%). 
However, the majority were tools that do not account for runtime deployment.

\section{Empirical Work}
\label{empirical}
In this section, we present the existing empirical work on hot fixes from the literature.

\subsection{Open Source Development}
\label{openSourceEmpirical}

A multitude of studies report empirical results from open-source projects.
Mockus et al.~\cite{Mockus2000} conduct a general study on open-source software development.
One of the conclusions they make is that in open-source systems, the response time to bugs reported by the customers is very quick.
In contrast to commercial settings, they found that bugs are patched as soon as they are reported.
Thus, an interesting takeaway from this work is that ``hot fixes'' might only be applicable to commercial systems that follow a stricter release schedule.

From a mobile application development perspective, Hassan et al.~\cite{Hassan2017} study the phenomenon of emergency updates in the Google Play store.
An emergency update in the context of this paper is an update published a short time after the previous update.
They discover that while these updates are not well documented, they often end up being a permanent change to the app and that they receive a lower ratio of negative reviews.
Moreover, they identify eight patterns for these types of updates.
The commonality in all of the discovered patterns is that these updates are usually due to simple development mistakes.
Shen et al.~\cite{Shen2017} also conduct an empirical study on the Google Play store to better understand how release planning can be optimized.
In analyzing the trends for user rating, the authors conclude that hot fixes are encouraged and do not harm the app as long as they strictly target fixing bugs as intended (not feature updates).

Within the context of video games, Lin et al.~\cite{Lin2017} study urgent updates which are either those that developers describe as hot fixes or updates outside of the planned release cycle for the purpose of fixing critical bugs that cannot wait until the next release.
While 80\% of the evaluated games in the study have urgent updates, they find that games that have a frequent update strategy are more likely to have a higher ratio of this kind of update.
The urgent update also does not necessarily always address an issue from the immediate previous update.
Truelove et al.~\cite{Truelove2021} 
investigate which kinds of bugs are most frequently targeted by hot fixes.
They find that crash bugs are the most severe, closely followed by Object Persistence and Triggered Event.

Illes-Seifert et al.~\cite{Illes-Seifert} consider hot fix the phase that makes up the first 5\% of the total time between two releases.
They find that the defect count of a file does not increase when it's modified in the context of a hot fix despite the usually inadequate testing of these changes.
Marconato et al.~\cite{Marconato2012} empirically analyze the hot fixing of security vulnerabilities. 
An interesting takeaway from this study is that developers are relatively very reactive to security bugs.
The average time for patching a vulnerability is around the 14-day mark. 
Moreover, the time between the discovery of a vulnerability and its patch release for operating systems decreases with the years.
Khomh et al.~\cite{Khomh2012} find that bugs are fixed faster when the length of the software release cycle is shorter, although a lower percentage of bugs are being fixed compared to longer release cycles.
Malone et al.~\cite{Malone2021} critique the manner in which software patches are released.
They find that only $1/4$ of security vulnerability patches are disclosed on the National Vulnerability Database.

Finally, Kolassa et al.~\cite{Kolassa2013} conduct an empirical analysis on the commit frequency in open-source software.
This is an analysis that can be used to inform configuration management activities which are essential to the hot fixing process.

\subsection{Commercial Development}


Savor et al.~\cite{Savor2016} study continuous deployment practices at Facebook and OANDA.
The authors in this paper use the number of hot fixes as a measure of software quality.
They find that increasing the number of deployments does not cause an increase in the number of hot fixes.
At Microsoft, Zhou et al.~\cite{Zhou2015} found that only 3.8\% of important customer issue reports are mitigated through hot fixes.

A study on Mandelbugs in real-world IT systems in production~\cite{Trivedi2011} creates estimates for several trends on hot fixing them based on the authors' experiences.
They disclose that a hot fix for a Mandelbug generally takes between a few minutes to three hours with the mean time for a hot fix being one hour.
Additionally, after manual analysis of a detected Mandelbug, it has a 0.2 probability of requiring a hot fix.
Finally, they estimate that the probability that such a hot fix would need just a reconfiguration and not a reboot is 0.9.
Anderson et al.~\cite{Anderson2015} touch on how development for hot fixing tends to happen on a separate branch and how this contributes to the large number of integration that happens during the initial stages of the release.

At CA Technologies, they find that the bug fixing time for different bugs is an uneven long tail distribution~\cite{Zhang2013}.
An analysis of the impact of various bug features on this bug fixing time is conducted in which they account for bug priority and bug severity.
In general, they find that bugs with the highest priority and highest severity are fixed faster. 
In the context of hot fixing, these are the type of bugs that hot fixing efforts would target.

Finally, Li and Long~\cite{Li2011} study the architectural degeneration of a commercial compiler system across two versions.
They find that architecture degenerates over time and that correlated components are the main cause for this degeneration.
As hot fixes are most commonly temporary workarounds for critical issues, we can assume that by their nature they contribute to this phenomenon.

\subsection{Human Research}
A few studies conduct surveys with users and system administrators which give insight into the practices and challenges in hot fixing software.
An important perspective on hot fixing that must be considered in user-facing systems is the end-user experience in the installation process itself once the hot fix is released.
Vaniea and Rashidi~\cite{Vaniea2016} surveyed $307$ users.
They found that users experience six stages when updating their system: awareness, decision to update, preparation, installation, troubleshooting, and post date.
Sarabi et al.~\cite{Sarabi2017} studied user behavior when it comes to software updating within the security domain by analyzing more than $400,000$ Windows machines.
They looked at the relationship between the vendors and the users taking into account the rate of updating, vendor patch deployment, and patch installation practices.

The second category of individuals that must be considered are the system administrators who manage the machines of organizations and their software updates.
To make sure that the released hot fixes actually get updated on the underlying infrastructure of an organization, we need to understand and cater to the needs of these individuals.
Li~et~al.~\cite{Li2019} found that system administrators go through five main stages: learning, deciding, preparing, deploying, and remedying.
This was found through surveying $102$ system administrators, $17$ out of them in-depth.
This study further identifies four pain points for system administrators: update information retrieval, update decision making, update deployment, and organizational culture that impedes regular update adoption.
Barrett et al.~\cite{Barrett2004} conducted field studies to further understand the problem-solving strategies of system administrators and conclude that available tooling does not support them in their practice, encouraging further work in this specific area.

\section{Techniques and Applications}
\label{techniquesAndApplications}
\begin{figure}
\centering\includegraphics[width=0.5\linewidth]{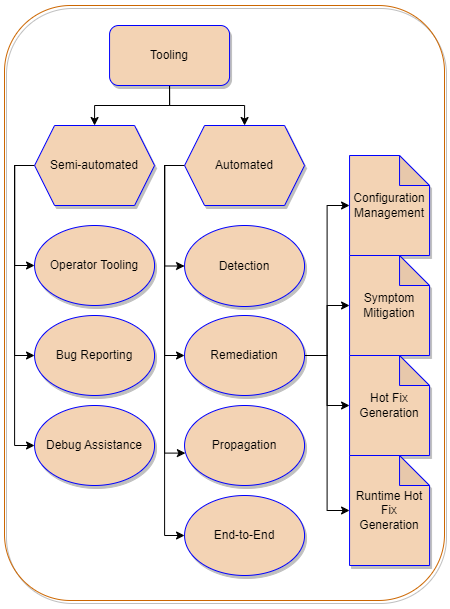}
 \caption{Taxonomy of existing work on tooling for hot fixing activities.}
 \label{fig:existingWorkTaxonomy}
\end{figure}

In this section, we detail techniques and applications that we found on hot fixing.
First, we present work on semi-automated tooling, i.e., helper tools for system administrators and software developers for hot fixing at different stages and granularities.
These tools do not fully automate any of the stages of the hot fixing process.
Instead, they aid the human effort in doing so through reporting, debugging, etc.
We then dive into fully automated tools that automate at least one of the stages of hot fixing: detecting the need for a hot fix, generating the hot fix, or deploying it into the target system.
We outline the taxonomy of existing work on hot fix tooling in Figure~\ref{fig:existingWorkTaxonomy}.


\subsection{Semi-automated Techniques}
\label{semiAutomated}
We first cover semi-automated techniques for hot fixes.
These are techniques that are implemented either as tooling for system administrators to help them in hot fixing activities, bug reporting techniques that ease the process of depicting and understanding the underlying critical issue, or tooling for assisting with debugging.

\subsubsection{System Administrator Tooling}

System administrators are often the individuals responsible for handling critical crises in enterprises.
They usually play the most critical role in mitigating time-sensitive system disruptions such as security attacks, performance degradation, and system unavailability.
While there has been abundant research on tooling for software developers, very little is known about the methods that the system administrators follow.
These individuals and the tools that they adopt are vital for hot fixing systems.

Bodik et al.~\cite{Bodik2006} propose two new tools for system administrators.
The first aids in tracing dependencies between the different components of the system and allows for tracking the overall health of the system.
The second monitors the actions of administrators to enable the identification of recurring problems and automate suggestions for their resolution.
In a continuation paper~\cite{Bodik2010}, they develop a technique for crisis classification based on summarising the data center's state in a representation called a \textit{fingerprint}.
The fingerprint is a summary of hundreds of performance metrics which can be used to identify whether a specific crisis has been seen before so that its known solution can be applied.

Another tool called TJOSConf~\cite{Wang2022} is a platform that assists with system management by automating system updates safely.
It interacts with services to verify the impact of an update and ensure that unexpected failures are recognized quickly. 
First-aid~\cite{Gao2009} is a runtime tool for diagnosing and patching memory management bugs and the memory objects that trigger them.
AUDIT~\cite{Luo2018} is a tool for troubleshooting recurring transient
errors.
Using lightweight triggers, it is able to identify the first occurrence of a problem and then rank the software methods according to their likelihood of being involved in the root cause of the problem.
Similar work was done by Yuan et al.~\cite{Yuan2006} which is based on statistical learning techniques that classify system call sequences.

\subsubsection{Bug reporting}

An efficient reporting process for critical issues in production code must be put in place to facilitate fast mitigation and hot fixing.
Several studies tackle this topic by proposing automated tooling for quicker incident diagnosis through optimised bug reporting processes.
Zhang et al.~\cite{Zhang2021} developed a tool called ``Onion'' that automatically locates incident-indicating logs.
Their solution uses log clustering techniques and has been used practically on the cloud system at Microsoft.
Log3C~\cite{He2018} is a tool that also uses log clustering for distinguishing relevant logs but with the addition of utilizing system KPIs (key performance indicators).
Similarly to Onion, Log3C has been successfully applied in the industry.

At Microsoft, the Windows Error Reporting (WER) has processed error reports collected from a billion machines over the course of 10 years~\cite{Glerum2009}.
This system classifies the collected reports into buckets based on similarity which allow for developer effort prioritization and more immediate, user-focused development.
Another technology at Microsoft is the Service Analysis Studio~\cite{Lou2013} which is a data-driven solution that tackles incident management from a software analytics perspective.\looseness=-1

Woodard et al.~\cite{Woodard2012} identify crises in complex distributed systems through automatic recognition of recurrences.
Using model-based clustering based on a Dirichlet process mixture, they are able to distinguish previous similar crises for which the root cause is known.

\subsubsection{Debug Assistance}

Although better bug-reporting methodologies and tooling can accelerate the process of identifying critical issues, additional assistance with debugging is often required to resolve them. 
The tool $\mbox{IDRA}\_\mbox{MR}$~\cite{Marra2020} is an online debugger for Map/Reduce applications that works by removing the debugging session to an external process. 
Wolverine~\cite{Verma2017} is an end-to-end debugging tool that allows for stepping through the execution of a program, the visualization of its states, and the synthesis of repair patches. 
Wolverine is able to integrate the synthesized patches into the running program without requiring a restart.
This tool can aid with the debugging process as well as suggest resolutions without the overhead of system downtime.

\subsection{Fully-automated Techniques}
Next, we review the applications of techniques that fully automate at least one of the stages for hot fixing.
This includes tooling for detection, remediation, and deployment.
Finally, we dedicate a section for tools we found that automate this pipeline end-to-end.

\subsubsection{Detection Tooling}

We begin with an overview of the tools that we found for detecting the need for a hot fix.
This includes tooling for detecting underlying functional and non-functional software issues.


Degradation of performance is often a problem that requires immediate attention through a hot fix.
Zhang et al.~\cite{Zhang2005} address the specific scenario of service level objectives violations.
They introduce an approach that identifies likely causes of performance problems within this context using several Bayesian network models that adapt to changing workloads.
Their approach is low cost which makes it feasible for practical application.
Fu et al.~\cite{Fu2012} also propose a solution for detecting critical performance issues.
Their technique uses system metric data mining to diagnose problems in online service systems.

Looking at problem diagnosis from the perspective of functional problems, Triage~\cite{triage} is an automated tool for onsite failure diagnosis.
This tool diagnoses a failure at the moment that it occurs and reports in detail the nature of the failure, the conditions that trigger it, as well as the fault propagation chain.
Triage mimics the steps a human takes in debugging a failure by employing several existing diagnosis techniques as well as a new one proposed by the authors called \textit{delta analysis}.
Khomh et al.~\cite{Khomh2011} address triaging crashes from the vantage point of prioritizing crash types.
This process helps direct development efforts toward critical crashes that might require hot fixing.
By grouping similar crash reports into crash types, they are able to use entropy region graphs to capture the distribution of their occurrences among system users.


Finally, from a security perspective, there have been several works that address diagnosing vulnerabilities.
Qin et al.~\cite{Qin2008} propose an automated mechanism \textit{Dataflow Analysis for Known Vulnerability Prevention System}.
Through tracing the vulnerability context and its spread path, they are able to detect the exploit and generate the vulnerability filter and its respective hot fix with minimal overhead. 
To protect vulnerable programs in the cloud, vPatcher~\cite{Zhang2015} examines network packets to detect vulnerable processes.
Araujo et al.~\cite{Araujo2018} present a cross-stack sensor framework for cyber security allowing for booby-trap insertion at multiple network layers.
FIBER~\cite{Zhang2018} is a tool that detects security vulnerabilities in software distributions by testing the presence of security patches.
FIBER does this by first analyzing open-source security patches, generating fine-grained signatures, and using these signatures to search against the target system.

Techniques that are based on prediction exist as well.
CRANE~\cite{Czerwonka2011} is a tool from Microsoft for failure prediction, change risk analysis, and test prioritization.
Its deployment helped Microsoft engineers identify tests that are likely to uncover problems.
Later research work at Microsoft on post-release defect prediction models utilizes various text execution metrics~\cite{Herzig2014}.
Finally, Shihab et al.~\cite{Shihab2011} propose prediction models that specifically focus on high-impact defects for customers and practitioners. These tend to be breakage defects and surprise defects respectively.
While they are somewhat successful in their mission, they conclude that more specialized models that take into account a defect's type instead of just its location are needed for practical adoption.

\subsubsection{Remediation Tooling}
\label{remediation}
Remediation for hot fixes takes many forms depending on the software issue.
Many critical issues are due to software configuration incompatibilities which can be fixed with reconfiguration rather than code-level patching.
Other issues require code changes but are extremely time-critical.
Thus, in these cases, instead of adding patches for the root cause which might be time consuming, workarounds might be sufficient so that the symptoms of the issue can be hidden as quickly as possible.
Finally, generating actual patches that address the underlying issues is sometimes needed.
Those can either be deployed offline or during runtime depending on the availability requirements of the target system.

\textit{2a) Configuration Fix}
Software configuration management is essential for achieving rapid changes to modern software systems.
Especially when developing hot fixes for software, the turnaround time must be fast. Configuration management makes it possible to monitor the changes made, package them, and make sure that they are error-free prior to delivery.
Karale et al.~\cite{Karale2016} propose a configuration management framework that automates the tasks of the configuration manager.
For hot fixing, their framework automates all the required steps for hot fix packaging which reduces the time required dramatically compared to a manual process.
To support applying configuration changes at runtime, Rasche et al. developed a new algorithm called ReDAC~\cite{Rasche2008} which ensures the consistency of application data while the reconfiguration happens.
ReDAC specifically supports distributed and multi-threaded component-based applications that have cyclic dependencies.
They specifically mention the case of hot fixing where dynamic reconfiguration is necessary to achieve high reliability and availability.

\textit{2b) Symptom Mitigation}
Often in the case of hot fixing, the bug is time-critical and there is not enough time to either identify the cause of the problem or generate a proper patch for it.
Thus, a common way to remediate critical bugs is through symptom mitigation.
The problematic side effect of the critical bug is removed, however, the root issue is not resolved.
This often comes at the cost of lost functionality or deteriorated performance.\looseness=-1

Qin et al. proposed Rx~\cite{Qin2005}, a technique implemented on top of Linux.
The idea behind it is that software failures can be mitigated by simply changing the environment in which the program executes.
The technique rolls the program back to the checkpoint where the program was in a stable state and then re-executes it in a modified environment.

Sweeper~\cite{Tucek2007} is a software security system that efficiently scans for suspicious requests.
After an attack is detected, Sweeper re-executes the suspicious code to apply analysis techniques.
Once the analysis is complete, Sweeper is then able to quickly recover the system and generate antibodies to prevent similar attacks from happening in the future.

Another security mitigation technique is Security Workarounds for Rapid Response (SWRRs) which has been implemented into a system called Talos~\cite{Huang2016}.
These techniques are designed to secure the system against vulnerabilities at the cost of lost functionality.
Unlike configuration workarounds which have a similar effect, SWRRs require minimal developer effort and knowledge of the system by utilizing existing error-handling code. 

\textit{2c) Offline Hot Fix Generation}
In this section, we expand on automated offline hot fix generation techniques.
We focus on techniques for generating hot fixes that require the system to be rebooted when the patch is applied.
In other words, these are techniques that generate hot fixes but do not take into account their runtime deployment into the system later.

Ding et al.~\cite{Ding2012}~\cite{Ding2014} mine historical issues and adapt them to suggest resolutions to similar software issues.
They are able to match issues by generating signatures for issues from their corresponding logs.
Their system was evaluated on a real system that compromises millions of users and was able to successfully provide resolutions that reduced the mean time to restore of the service.
Pozo et al.~\cite{Pozo2019} propose a protocol that hot fixes link failures in time-triggered schedules within a few milliseconds. 

Various techniques target offline hot fix generation for software vulnerabilities.
MacHiry et al.~\cite{MacHiry2020} argue that most security patches are safe patches meaning that they can be applied without disrupting the functionality of the program and thus require no testing.
They find that most patches in the CVE database are indeed safe patches which they hope will encourage project maintainers to apply them without a delay.
ShieldGen~\cite{Cui2007} is a technique that is able to generate a patch for a zero-day attack given its instance.
Subsequently, it is able to generate additional potential attack instances to determine whether given the patch the system can still be exploited.
AutoPaG~\cite{Lin2007} targets out-of-bound vulnerabilities specifically.
It catches the violation, and based on the data flow analysis is able to identify the root cause and generate a patch automatically.
The most impressive part of this work is that the vulnerability patch can be generated within seconds.
Tunde-Onadele et al.~\cite{TowardsJIT}~\cite{Olufogorehan2020} propose tooling for security attack containment that performs both exploit identification and vulnerability patching for containerized applications specifically.
They report accurate detection and classification for 81\% of attacks as well as an 84\% reduction in patching overhead.
For Android, we found AppSealer~\cite{Zhang2014} which automates fix generation for component hijacking vulnerabilities.
Finally, Aurisch et al.~\cite{Aurisch2018} propose using Mobile Agents for vulnerability and patch management.

\textit{2d) Runtime Hot Fix Generation}
In this section, we focus on hot fix generation techniques that account for the ability of the generated hot fix to be integrated into the system during runtime.
Generating this specific type of hot fix has been researched for multiple specific use cases.

AutoPatch~\cite{Salehi2022} is the first automated technique for hot fixing embedded devices with high availability on the fly.
Within memory management, for example, Exterminator~\cite{Novark2007} uses randomization to find memory errors and derive runtime patches for them.
First aid~\cite{Gao2009} is also a system tailored for treating memory management bugs.
It came after Exterminator to reduce the space and time overhead and allow for scaling. 
By rolling the program back to previous checkpoints, it is able to diagnose memory bugs.
Following the diagnosis, First-aid then generates patches that prevent the memory bug and applies them during runtime.

Hot fixing is perceived to have a high risk of leaving the system in an inconsistent state.
Katana~\cite{Ramaswamy2010} is a tool for hot fixing ELF binaries that aims to reduce this risk.
By introducing a new file format called a Patch Object, they are able to provide more information about the structure and implications of patches.
Another attempt to encourage users to use runtime patching is binary quilting~\cite{Saieva2020}.
This technique creates an entirely new reusable binary.
With this, users can apply the minimum patch required without the unwanted side effects.
For string interpolation vulnerabilities, DEXTERJS~\cite{Parameshwaran2015} is a low-overhead technique that automatically generates patches to place on vulnerable sites.
ProbeGuard~\cite{Bhat2019} balances performance and security by hot fixing more powerful defenses when probing attacks occur.
Xu et al.~\cite{Xu2020Source} propose a source code and binary level vulnerability detection and patching framework.
It learns from the source code the function inputs that trigger the vulnerability and builds a filter to block them.

As previously explained, there is great value in hot fixing during runtime.
However, many of the official patches on CVE do not allow for it.
VULMET~\cite{Xu2020Automatic} is a tool that learns from existing official patches and generates hot fixes that can be deployed during runtime by using  weakest precondition reasoning.
Similarly, EMBROIDERY~\cite{Zhang2017} transforms official CVE patches into hot fixes for a broad spectrum of Android devices.
The main contribution of this tool is that it can be used to maintain obsolete Android systems and devices that often do not receive patches.
A major hurdle with quick hot fixing for mobile devices is that Android partners require lengthy compatibility testing for the patches.
InstaGuard~\cite{Chen2018} tackles this problem by bypassing the testing requirement through avoiding the injection of new code and relying on rule generation instead.

Finally, also based on binary runtime injection techniques, band-aid patching~\cite{Sidiroglou2007} is a technique for testing hot fixes so that their deployment can be accelerated.

\subsubsection{Deployment Tooling}


An important consideration for hot fixing critical bugs is how the hot fix can ultimately be deployed efficiently after it is developed.
There have been a number of works that tackle this problem from different perspectives.
The techniques differ greatly depending on the target system into which the hot fix will be deployed i.e. Android systems, IoT infrastructures, cloud platforms, etc.
In this section, we cover papers on the deployment of hot fixes into different types of production systems.

To allow software fixes and even features to be deployed more quickly, Van der Storm~\cite{Van2005}~\cite{storm2007} proposes a solution for automating the delivery of components in component-based software.
When deploying software fixes to systems in production, the binary is modified and thus a restart is often required.
This leads to downtime of the system which is often disruptive, especially in user-facing and mission-critical systems.
There has been a lot of research into how patches can be integrated during run-time without causing the system to become unavailable as a result of the update.

As explained in Section \ref{terminologyEvolution}, this activity is often referred to as hot patching.
As discussed, since a survey on hot patching exists, our scope only includes work that hot patches time-critical issues, according to our definition of hot fix.
Payer et al.~\cite{Payer2013} call this a problem of ``as soon as possible'' (ASAP) repair. 
They investigate the feasibility of runtime patching by investigating whether patches for critical bugs from the Apache web server can be dynamically applied.
They find that dynamic update mechanisms are feasible and highly effective.
Katana~\cite{Ramaswamy2010} is one such dynamic update mechanism.
Katana specifically hot patches ELF binaries by creating \textit{patch objects} which the user can apply to running processes.
The authors of this paper also address an important concern that many owners have regarding hot patching.
As this type of updating mechanism is less common, these techniques are often viewed as at risk of leaving the system in an inconsistent state. 
Russinovich et al.~\cite{Russinovich2021} present an optimization to live migrations of virtual machines that does not require turn space, requires minimal CPU, and no network while preserving VM state and causing minimal VM blackout.

Users are sometimes reluctant to update their system once a hot fix has been deployed due to the risk of software releases becoming incompatible and breaking their system.
Saieva et al.~\cite{Saieva2022} propose alleviating this friction through a technique called \textit{Binary Patch Decomposition}.
This technique provides the users with extra context around the distributed fixes which allows them to select and integrate the compatible pieces of the update.
To increase the robustness of upgrades but in the context of rolling upgrades on cloud platforms is an approach named $R^2C$~\cite{Sun2018}.
This approach offers early error detection of rolling upgrades as well as risk and time completion predictions.

Several works address the deployment of hot fixes for Android systems specifically.
InstaGuard~\cite{Chen2018} is a hot patching approach for Android that bypasses the slow compatibility testing processes of Android device partners to facilitate immediate patching  for critical security vulnerabilities.
InstaGuard makes this possible by avoiding the injection of new code in the patch and enforcing instantly updatable rules instead.
Ford et al.~\cite{Ford2018} discuss two tools that tackle the same problem as Instaguard by bypassing the traditional mobile app patching lifecycle.
They find that while both of these tools do enable quick updates, they expose the users to many security vulnerabilities.
PatchDroid~\cite{Mulliner2013} is a tool that distributes and applies Android security patches.
It enables safe in-memory patching in a scalable way such that a patch can be written once and deployed to all affected versions.
Socio-Temporal Opportunistic Patching (STOP)~\cite{Tang2012} also enables the delivery of security hot fixes to mobile devices.
It is a two-tier system that collects co-location data of mobile devices then targets the delivery of hot fixes to a subset of these devices.
From there, the patch can be spread by these devices opportunistically.
LEONORE~\cite{Vogler2015}~\cite{Vogler2016} is a large-scale IoT deployment framework that allows for the provisioning of components on resource-constrained edge devices.
Araujo et al.~\cite{Araujo2020} propose a patch management model for rapid deployment of security patches during runtime.
Their approach includes patch testing and recovery in the case of an incompatible patch.

In complete contrast with the idea of patching during runtime, Candea et al.~\cite{Candea2001} highlight the importance of recursive restartability.
In other words, the ability of the system to handle restarts at multiple levels.
This is because many nondeterministic bugs do not necessarily need to be patched.
They can be solved by simply rebooting.
In their paper, they highlight the required properties for recursive restartability and outline steps for beginning to adopt this in software systems.
In a continuation paper~\cite{Candea2004}, they discuss microrebooting.
The idea is to recover the faulty application components without affecting the rest of the application.

\subsubsection{End-to-End Tooling} 

Our literature review concludes with a few works that outline end-to-end approaches to automate the detection, remediation, and deployment of hot fixes for critical software issues.

The earliest work we could find that tackles this is Huang et al. in 2005~\cite{Huang2005}.
The authors propose an automated hot fixing framework in which they are able to reason about the cause of a fault, apply simple remediation patching, and do this during runtime without affecting the system's availability.
They evaluate their technique on a small-scale within the domain of web-based applications.

Dataflow Analysis for Known Vulnerability Prevention System~\cite{Qin2008} is an end-to-end technique for protecting a target system at risk from security attacks.
It detects attacks through dataflow analysis, examines the spread of the attack, and generates an appropriate vulnerability filter and hot fix which can be deployed at runtime.

Gomez et al.~\cite{Gomez2015} present a new vision for app stores. 
This new generation of app stores aims to automate several functionalities in mobile app maintenance to improve their quality and reduce human intervention.
By utilizing user reviews, ratings, execution traces, and crash reports they aim to monitor and analyze the app and then automatically be able to generate, validate, and deliver patches.
This new generation of app stores would work in a feedback loop model such that the newly delivered patches would result in new input data that they can then use to produce even more patches.
They begin to realize their vision for Android devices in a continuation paper~\cite{Gomez2017}.
The framework was tested minimally on a single app in which they were able to patch a user-reported crash.

Itzal~\cite{Durieux2017} is an automated software technique that generates patches directly in the production environment.
The novelty of this tool is that it removes the requirement of having a failing test case.
Instead, it accesses the system state at the point of failure to conduct regression testing and patch search.
The paper discusses a proof of concept prototype implementation which was evaluated on null dereference failures specifically.
Such tooling can accelerate the process of generating hot fixes as it not only automates the patch generation process but is designed to work directly in production.

Wolverine~\cite{Verma2017} is an interesting tool that addresses the three phases of detection, repair, and deployment.
However, it does this within the context of debugging sessions.
By allowing the developer to step through the program states and visualize them, Wolverine aids with detection.
It then implements a repair algorithm that synthesizes patches.
To avoid having to abort the debug session, it allows for hot fixing suggested patches during runtime.
It has been evaluated on somewhat small programs (student submissions and programs that implement known data structures).

AFix is another tool that automates this whole process~\cite{Jin2011}.
However, it only targets a single common type of bug: single-variable atomicity violations.
It is first able to detect these bugs from bug reports.
Using static analysis, AFix constructs suitable patches for the detected bugs.
Finally, it attempts to combine multiple patches to improve performance and readability.
It provides customized testing for each patch for validation.
Sidiroglou et al.~\cite{Sidiroglou2005} also only look at a specific set of bugs, those that are recurring within the system.
Using an instruction-level emulator before instruction execution, they are able to check for recurring faults and recover a safe control flow for the program.

Finally, ClearView~\cite{Perkins2009} is a five-step automated system for patching systems with high availability requirements.
It learns invariants of the program behavior, monitors for failures, identifies failures through invariant violations, generates patches to uphold the invariants by changing the state or flow of control, and observes the execution after patches are applied to select the most successful patch.
As for evaluation, an external Red Team generated 10 code injection exploits to attack an application protected by ClearView.
ClearView was able to block all of the attacks.
It was able to generate patches that correct the behavior of the system under attack for 7 out of the 10 cases.

These tools are inspiring as they demonstrate the feasibility of automation in this domain and the implications that this can have on software development.
However, they remain quite limited in their application and thus hint that additional research and development in this area is needed.

\section{Research Agenda and Open Challenges}
On the basis of this literature review, we present our recommendations to the software engineering research community through the following open challenges (see Table~\ref{tab:openChallenges}):
\begin{table*}
  \caption{Open challenges in hot fixing.} 
  \centering
  \label{tab:openChallenges}
  \begin{tabular}{llll}
    \toprule
    \textbf{Open Challenge} & \textbf{Priority} & \textbf{Impact}\\ 
    \midrule
    The Hot Fixing Vocabulary Challenge & 1 & Unified definitions will ensure that future empirical work,\\
     & &  benchmarks, tooling, etc address the correct problem.\\
    \midrule
    The Hot Fixing Benchmarking Challenge & 2 & 
    Specialized benchmarks are needed to understand the\\
     & & current state-of-the-art and eventually to evaluate the \\
     & & performance of hot fixing tooling that will be built.\\
    \midrule
    The Hot Fixing Taxonomy Challenge & 3 & Outlining a taxonomy for what issues hot fixing targets\\
    & & will ensure that tooling is sufficient and complete.\\
    \midrule
     The Hot Fixing Industrial Practice Challenge
 & 4 & Understanding the current industrial practices will aid\\
     & & in understanding the pain points and help in building\\
     & & useful tooling. Additionally, candid conversations with\\
     & & developers who work on hot fixing are essential to \\
     & & creating a body of work that is useful to them. \\
    \midrule
    The Hot Fixing Distribution Challenge & 4 & Gaining a deeper understanding of hot fixing patterns \\
    & & will help systematize this activity.\\
    \midrule
    The Hot Fixing Impact Challenge & 4 & Critical issues are the targets of hot fixes. Thus, studying\\
    & & their impact is integral to developing successful hot\\
    & & fixing tooling.\\
    \midrule
    The Hot Fixing Tooling Challenge & 5 & There are many directions for future work in developing\\  
    & &  hot fixing tooling. Such tooling can make hot fixes faster \\
    & & to develop and deploy and lower its risk on the \\
    & & production environment.\\
    \midrule
    The Hot Fixing Predictability Challenge & 6 & Accelerate the process of identifying critical issues and\\
    & & thus accelerating their hot fixing process.\\
  \bottomrule
\end{tabular}
\end{table*}

\begin{enumerate}
    \item \textbf{The Hot Fixing Vocabulary Challenge}: Using an unified terminology to aid understanding of the state-of-the-art in the area and ease building upon existing work.\looseness=-1
    \item \textbf{The Hot Fixing Benchmarking Challenge}: Building a benchmark of hot fixes to aid research comparisons and ease progress in the area. 
    \item \textbf{The Hot Fixing Taxonomy Challenge}: Creating a taxonomy for critical bug classes and remediation strategies that target them to drive research toward more advanced remediation tooling. 
    \item \textbf{The Hot Fixing Industrial Practice Challenge}: Conducting more empirical studies on industrial practices in terms of automated and manual strategies for detecting, remediating, and deploying hot fixes. This can also include conducting surveys on hot fixing practices with developers.\looseness=-1
    \item \textbf{The Hot Fixing Distribution Challenge}: Providing details on the cost of hot fixing in software enterprises, the average frequency of hot fix releases, and a structured taxonomy on the critical bug types that they target to aid future research and understand the hot fixing cost in practice.\looseness=-1
    \item \textbf{The Hot Fixing Impact Challenge}: Measuring the impact of a critical issue and predicting the best remediation strategy for it as a first step towards automating hot fixing. This includes automation for quantifying the time allotted to find the fix, the issue's direct impact on users, and the desired type of resolution. Existing work that tackles some of the sub-tasks of this mission can be combined to fully automate this process.
    \item \textbf{The Hot Fixing Tooling Challenge}: Developing specialized tooling for hot fixing for all of the phases that we identified in this survey: identifying critical issues, remediating these issues, and finally deploying the hot fixes as quickly as possible.
    Automation and the creation of end-to-end specialized tooling here opens up many directions for future work.
    Another tool we have not seen is a tool that measures the impact of a hot fix post-deployment.\looseness=-1
    \item \textbf{The Hot Fixing Predictability Challenge}: Exploring the power of existing defect and vulnerability prediction tooling on the set of bugs that have been hot-fixed, to understand their ability to detect such bugs.\looseness=-1
\end{enumerate}

While these open challenges might seem as basic recommendations, they are yet to be explored within the context of hot fixing.
This makes these challenges even more paramount as we find that many of the building blocks for facilitating advancements in research within the area remain untapped.
We hope that these recommendations highlight the need for this foundational research to the research community.

\section{Discussion and Reflections}
\label{discussion}
In this section we expand on the aforementioned challenges 
through a detailed discussion and reflection on the existing body of work.
We also point out directions for future work. 


\subsubsection{Terminology}
It is very evident from this review that the terminology used in different papers is sometimes conflicting. This has created a disjointed body of research.
A first step in advancing the research on hot fixing is unifying the terminology and reflecting on how all of the papers that address this topic fit together -- our aim with this comprehensive review.
Moreover, some of the existing work included in our review is applicable to hot fixing, since it addresses the treatment of critical bugs, even though it does not explicitly mention any hot fixing terminology.
Thus, we urge the community to not only produce more research directed at hot fixing but to pay careful attention to consistency of terminology so as to streamline the acquisition of collective knowledge.

\subsubsection{Benchmarks}
From here, an essential requirement for driving research on a specific topic forward is reliable benchmarks.
In our review, we were not able to find any such large-scale benchmarks on hot fixing.
We believe that having a collection of real-world hot fixing instances can help in better understanding this software engineering activity as observed in production.
As explained in section~\ref{remediation}, we were able to identify multiple remediation techniques depending on the type of critical issue.
Thus, a larger benchmark is needed here to realistically capture the different hot fix types depending on the granularity (e.g. build, configuration, source code) as well as the robustness of the fix (e.g. workaround, root cause fix).
Such a benchmark can help in determining the effectiveness of existing tooling so that research efforts can be directed more productively.
Tens of millions of public repositories exist on GitHub\cite{github} which can be utilized in creating it.
Taking this a step further, it would be valuable to collaborate with industrial stakeholders to create benchmarks from larger-scale products.
As the size of the product grows, prioritizing the issues and detecting the most critical ones inevitably becomes a more difficult task.
Generating a hot fix for these issues becomes more complex as there are likely more dependencies that would need to be taken into account.
Thus, we hypothesize that the scale of the target system would have some effect on what constitutes a hot fix and what the classes of hot fixes would be.


\subsubsection{Empirical Work}

 In open source empirical research on hot fixing, mobile applications and video games have been the most studied assessing the types of updates and the bugs that they target (Section~\ref{openSourceEmpirical}).
Assessing these properties in different domains, such as security, cloud, web, and machine learning models, can be further explored in the literature. We believe that this is essential for drawing wider conclusions from the empirical results.

Multiple smaller-scale studies have been conducted in industrial settings.
There has been industry-partnered research in which hot fixing is mentioned but is not the central topic of the work~\cite{Zhang2013}.
In the future, we hope to see larger-scale studies on hot fixing efforts in real-world applications.
Moreover, we found that there has not been large-scale industrial adoption for hot fix automation tooling as of yet.
We hope that future research on hot fixing creates a positive feedback loop in industrial settings.

The literature lacks detail on the cost of hot fixing in software enterprises, the average frequency of hot fix releases, and a structured taxonomy on the critical bug types that they target. This quantitative information is important for incentivizing companies to invest in hot fix automation tooling and further understanding the level of automation that is actually required to optimize productivity and user satisfaction.

Finally, while surveys and interviews have been conducted with end users~\cite{Vaniea2016} and system administrators~\cite{Li2019} there is still a gap in conducting such human empirical work with the software developers.
Understanding patterns in hot fixing activities from their perspectives can bring the required insight for understanding the prevention of bugs targeted by hot fixing.

\subsubsection{Semi-Automated Tooling}



Semi-automated tooling whose purpose is to aid system administrators with treating critical crises needs to mimic the user journey of the administrator.
In this section, we break down the user journey into its different stages and address possible future research directions given existing work as presented in Section~\ref{semiAutomated}.

The administrator needs to get notified of the crisis in a user-friendly way to increase productivity.
Techniques for localizing logs that demonstrate the incident via log clustering~\cite{Zhang2021} and system KPIs~\cite{He2018} make this possible.
They then need to know the impact of the crisis.
This includes quantifying the number of users that it affects, whether backups exist, the criticality of the features/projects affected, and so on.
This will inform the amount of time that they have to resolve the crisis.
We found tooling for tracing dependencies~\cite{Bodik2006}, summarizing a data center's state using system metrics~\cite{Bodik2010}, and effort prioritization given a collection of error reports~\cite{Glerum2009}.
We believe that there remains potential here in taking this a step further.
We envision a tool that brings these tools together.
Given the dependency tracing, the system's current state, and effort prioritization, we need to automate quantifying the time allotted to find the fix, the issue's direct impact on users, and the desired type of resolution (compromising features, finding the root cause solution, securing the system, etc.), assigning the task to the most qualified administrator for the task, as well as providing the required data for the following step.

From here, they will begin resolving the crisis.
The resolution will be affected by the criticality of the crisis and the time allotted to resolve it.
As such, for crises that need immediate resolutions, workarounds will be implemented whereas for crises that allow for more time, a true resolution might be achieved.
Automated tools that detect recurring crises~\cite{Woodard2012}~\cite{Bodik2006} and suggest resolutions for them can greatly help in increasing the turnaround time as well as increase productivity so that they do not have to resolve the same issue multiple times.
Automated repair techniques exist~\cite{surveyLeGoues}, however, none target the needs of system administrators directly.
In the future, we would like to see a clear taxonomy of the types of bugs that are observed by these individuals to help drive research forward.
We expect that the different classes of critical bugs require vastly different remediation strategies.
Given the taxonomy, we will hopefully begin to see more specialized automated repair tooling for system administrators.

Once a resolution has been reached, it must be deployed as quickly as possible ideally without causing any downtime to the target system.
We were able to find a tool for measuring the impact of an update for better system update safety~\cite{Wang2022}.
From here, verifying that the resolution did indeed resolve the crisis is required.
There is a gap here in the literature.
We envision an automated tool here dedicated to managing recently deployed resolutions.
Specifically in tooling for verifying the satisfaction with the deployed fix, measuring its impact post-deployment, and specifying whether it needs to be improved at some point into a permanent fix, when this fix would need to take place, and who would be responsible for undertaking it. 

The Microsoft Service Analysis Studio~\cite{Lou2013} is the closest tool we found in this domain to more advanced system administrator tooling.
However, this paper was published over $10$ years ago.
Since the internal process for software engineering tasks are ever-changing and advancing with the growing size and complexity of systems, we believe bringing such a study up to date can be extremely valuable.
Thus, we encourage the community to publish research in this area so that we can begin to see a new generation of such tools targeted specifically at time-critical software issues.

\subsubsection{Detection Tooling}



We were able to find detection of critical software issues in the context of performance degradation, functional defects, and security vulnerabilities.
We specifically only wanted to include tools that mention identifying \textbf{critical} issues.
Our search concluded with only $26$ papers,
suggesting that more research in this area is needed.

None of the studies that we found in the context of detecting critical issues specifically target prediction.
There is abundant research on vulnerability~\cite{Shen2020} and defect prediction~\cite{Thota2020}.
However, through this review, we were not able to see a clear correlation between those works and hot fixing.
We believe that there is untapped potential in the applicability of these prediction techniques for this context.

\subsubsection{Remediation Tooling}






For the purpose of hot fixing, there is no one-size-fits-all solution.
A fully optimized remediation framework for critical software issues must sit on top of all of the remediation strategies that we mention in Section \ref{remediation}: reconfiguration, symptom mitigation (workarounds), as well as offline and online hot fix generation.
Depending on the nature of the critical issue, the correct remediation strategy must be applied.
This will depend on the type of bug, its impact, the complexity of the resolution, the availability requirements of the system that the issue resides in, and so on.
The major gap in the literature that we see here is not necessarily in advancing each of the remediation strategies.
We believe that the first step that needs to be tackled here is deepening the collective understanding of what strategy needs to be applied and when.
This can then be unified in a switch case-like system that automates the assignment of each critical issue to the correct strategy and launches that corresponding automated remediation technique.

\subsubsection{Deployment Tooling}
\label{propagationDiscussion}

Despite the abundant existing research on runtime patching of software, these techniques are not often applied in practice. 
One of the major reasons is that software developers and maintainers view them as high-risk for breaking the target systems.
Thus, we think that more human-studies are required to understand the needs of these individuals in order to build tools that they can put their trust in.
More empirical research on the success of these techniques can help in shifting their perspective as well.
Finally, simulation environments can help validate the runtime patch before deployment.

\subsubsection{End-to-End Tooling}


One of the main takeaways we wish to leave you with in this survey is that end-to-end tooling for automating hot fixes is lacking.
We were only able to find four tools that address all three phases of treating critical software issues.
However, these tools are either within a specific domain (i.e., mobile~\cite{Gomez2015}, web apps~\cite{Huang2005}), or target a specific kind of critical software issue (i.e. single variable atomicity violations~\cite{Jin2011}, online debugging~\cite{Verma2017}).
We believe that there is a big gap in the literature here.
This task of hot fixing critical software bugs is needed in every system in production.
Directed research on this specific task and how it can be partially/fully automated will have an immense impact on the software engineering community and a wide variety of applications in industrial settings.
We encourage researchers to further expand the existing research into different use cases and domains.

As explained in the survey methodology (Section \ref{methodology}), our scope encapsulates work that targets hot fixing activities specifically.
However, much bug detection, remediation, and deployment tooling exists that are not targeted at critical software issues or hot fixing.
Comparative studies that analyze the efficacy and efficiency of this more general tooling but in our use case, as opposed to hot fixing-specific tooling, would be useful to better understand the current state of the art.

\section{Threats to Validity}
\label{threats}
The relevance filtering of the papers in the search was conducted by one author of this paper.
As this can be subject to judgment bias, a second author of this paper acted as a second judge on a sample of the papers to measure agreement.
A $10\%$ random sample of the irrelevant papers and $10\%$ random sample of the relevant ones were selected and shuffled.
The second judge then manually assessed and filtered the papers based on the scope of the literature review.

The second author's judgment matched that of the first judge in all but two instances.
There were two papers that were included in the scope of the survey that the second author deemed as irrelevant.
These were papers that did not explicitly mention hot fixing or addressing critical software issues.
Instead, they present transferable techniques that can be applicable to the problem as they automate the detection of bugs and the deployment of patches into the end system.
Thus, we conclude that the judgment of the author who collected the papers does not miss any relevant work.
However, they were more lenient with their inclusion criteria which might have resulted in more papers getting included in this survey.

In our primary search, we use two keywords: \textit{hot fix} and \textit{hot patch}.
We attempted to look for synonyms of these keywords in the IEEE thesaurus~\cite{IEEEThesaurus} but were unable to find any.
Thus for due diligence, we conducted a primary search on keywords we see as related to the scope of this literature review.
We conducted this search on one search engine only to gauge relevance which was IEEE Xplore.
We experimented with the following keywords: \textit{critical bug fix}, \textit{critical bug patch}, \textit{critical fix}, and \textit{critical patch}.
We found that the number of search results returned was very small (3, 0, 14, and 51 respectively).
In addition, only 5 papers were relevant to our scope and out of these 5 papers none were key papers that would add substantial additional knowledge to the contents of this review.

Finally, the categorization of the papers is subject to bias.
We mitigate this by following the thematic analysis~\cite{Braun2012} procedure used in qualitative research to conduct this in a more structured manner.

\section{Conclusions}
\label{conclusions}


We have presented an overview of existing work on hot fixing for software systems.
Hot fixing is a fundamental software engineering task for systems in production.
However, we found that the number of publications on the topic remains limited and the terminology is yet to be streamlined.
We believe that hot fixing both in industry and research has the potential to become a more systematic and established software engineering activity, and less of an afterthought than it is viewed today.
With this paper, we wish to ignite more research in the area, especially toward automation.
In the future, we hope to see community effort in building benchmarks and conducting both quantitative and qualitative empirical studies on current practices.
Following a deeper understanding of the state of the art and current challenges, we then hope to start seeing the process become more automated.

\section*{Acknowledgments and Copyright}
We would like to thank those authors who provided comments and feedback on earlier drafts of this paper. We would also like to thank Prof. Mark Harman for his valuable input at various stages of this work. 
This work was supported by the UKRI EPSRC Fellowship EP/P023991/1 and the ERC Advanced Grant No.741278.
For the purpose of open access, the authors have applied a Creative Commons Attribution (CC BY) license to any accepted manuscript version arising.

\bibliographystyle{acm}
\bibliography{main}

\begin{thebibliography}{100}

\bibitem{acm}
{ACM Digital Library}.
\newblock \url{https://dl.acm.org/}.
\newblock Accessed: 2023-11-17.

\bibitem{dblp}
dblp: computer science bibliography.
\newblock \url{https://dblp.org/}.
\newblock Accessed: 2023-11-17.

\bibitem{github}
Github: Let’s build from here.
\newblock \url{https://github.com/}.
\newblock Accessed: 2023-11-17.

\bibitem{ieee}
{IEEE Xplore}.
\newblock \url{https://ieeexplore.ieee.org/Xplore/home.jsp}.
\newblock Accessed: 2023-11-17.

\bibitem{sciencedirect}
{ScienceDirect}.
\newblock \url{https://www.sciencedirect.com/}.
\newblock Access: 2023-11-17.

\bibitem{Agarwal2014}
{\sc Agarwal, A., and Garg, N.~K.}
\newblock Effective test strategy model for ensuring ftr of hot-fix.
\newblock {\em ICROIT 2014 - Proceedings of the 2014 International Conference on Reliability, Optimization and Information Technology\/} (2014), 40--43.

\bibitem{testingChallengesFBISSTA2019}
{\sc Alshahwan, N., Ciancone, A., Harman, M., Jia, Y., Mao, K., Marginean, A., Mols, A., Peleg, H., Sarro, F., and Zorin, I.}
\newblock Some challenges for software testing research (invited talk paper).
\newblock In {\em Proceedings of the 28th ACM SIGSOFT International Symposium on Software Testing and Analysis\/} (New York, NY, USA, 2019), ISSTA 2019, Association for Computing Machinery, p.~1–3.

\bibitem{STchallengesICST23}
{\sc Alshahwan, N., Harman, M., and Marginean, A.}
\newblock Software testing research challenges: An industrial perspective.
\newblock In {\em 2023 IEEE Conference on Software Testing, Verification and Validation (ICST)\/} (2023), pp.~1--10.

\bibitem{Anderson2015}
{\sc Anderson, J., Salem, S., and Do, H.}
\newblock Striving for failure: An industrial case study about test failure prediction.
\newblock {\em ICSE\/} (2015).

\bibitem{Araujo2020}
{\sc Araujo, F., and Taylor, T.}
\newblock Improving cybersecurity hygiene through jit patching.
\newblock {\em ESEC/FSE\/} (11 2020), 1421--1432.

\bibitem{Araujo2018}
{\sc Araujo, F., Taylor, T., Zhang, J., and Stoecklin, M.~P.}
\newblock Cross-stack threat sensing for cyber security and resilience.
\newblock {\em Proceedings - 48th Annual IEEE/IFIP Int. Conf. on Dependable Systems and Networks Workshops, DSN-W 2018\/} (7 2018), 18--21.

\bibitem{Aurisch2018}
{\sc Aurisch, T., and Jacke, A.}
\newblock Handling vulnerabilities with mobile agents in order to consider the delay and disruption tolerant characteristic of military networks.
\newblock {\em Int. Conf. on Military Communications and Information Systems\/} (6 2018), 1--7.

\bibitem{Barrett2004}
{\sc Barrett, R., Kandogan, E., Maglio, P.~P., Haber, E.~M., Takayama, L.~A., and Prabaker, M.}
\newblock Field studies of computer system administrators: analysis of system management tools and practices.
\newblock {\em CSCW '04: Proceedings of the 2004 ACM Conf. on Computer supported cooperative work\/} (11 2004), 388--395.

\bibitem{Bhat2019}
{\sc Bhat, K., Kouwe, E. V.~D., Bos, H., and Giuffrida, C.}
\newblock Probeguard: Mitigating probing attacks through reactive program transformations.
\newblock {\em Int. Conf. on Architectural Support for Programming Languages and Operating Systems - ASPLOS\/} (4 2019), 545--558.

\bibitem{Bodik2006}
{\sc Bodík, P., Fox, A., Jordan, M.~I., Patterson, D., Banerjee, A., Jagannathan, R., Su, T., Tenginakai, S., Turner, B., Ingalls, J., Lab, R., Berkeley, U.~C., and University, S.}
\newblock Advanced tools for operators at amazon.com.
\newblock {\em Hot Topics in Autonomic Computing (HotAC)\/} (2006).

\bibitem{Bodik2010}
{\sc Bodík, P., Goldszmidt, M., Fox, A., Woodard, D.~B., and Andersen, H.}
\newblock Fingerprinting the datacenter: Automated classification of performance crises.
\newblock {\em EuroSys'10 - Proceedings of the EuroSys 2010 Conf.\/} (2010), 111--124.

\bibitem{Braun2012}
{\sc Braun, V., and Clarke, V.}
\newblock Thematic analysis.
\newblock {\em APA handbook of research methods in psychology, Vol 2: Research designs: Quantitative, qualitative, neuropsychological, and biological.\/} (3 2012), 57--71.

\bibitem{Candea2001}
{\sc Candea, G., and Fox, A.}
\newblock Recursive restartability: Turning the reboot sledgehammer into a scalpel.
\newblock {\em Proceedings of the Workshop on Hot Topics in Operating Systems - HOTOS\/} (2001), 125--130.

\bibitem{Candea2004}
{\sc Candea, G., Kawamoto, S., Fujiki, Y., Friedman, G., and Fox, A.}
\newblock Microreboot-a technique for cheap recovery.

\bibitem{Chen2018}
{\sc Chen, Y., Li, Y., Lu, L., Lin, Y.-H., Vijayakumar, H., Wang, Z., and Ou, X.}
\newblock Instaguard: Instantly deployable hot-patches for vulnerable system programs on android.
\newblock {\em Network and Distributed System Security Symposium\/} (2018).

\bibitem{Cui2007}
{\sc Cui, W., Peinado, M., Wang, H.~J., and Locasto, M.~E.}
\newblock Shieldgen: Automatic data patch generation for unknown vulnerabilities with informed probing.
\newblock {\em Proceedings - IEEE Symposium on Security and Privacy\/} (2007), 252--266.

\bibitem{Czerwonka2011}
{\sc Czerwonka, J., Das, R., Nagappan, N., Tarvo, A., and Teterev, A.}
\newblock Crane: Failure prediction, change analysis and test prioritization in practice - experiences from windows.
\newblock {\em Proceedings - 4th IEEE Int. Conf. on Software Testing, Verification, and Validation, ICST 2011\/} (2011), 357--366.

\bibitem{criticalBugs}
{\sc D'Ambros, M., Lanza, M., and Pinzger, M.}
\newblock A bug's life visualizing a bug database.
\newblock {\em Int. Workshop on Visualizing Software for Understanding and Analysis\/} (2007), 113--120.

\bibitem{Dijkstra1969}
{\sc Dijkstra, E.~W.}
\newblock Structured programming.

\bibitem{Ding2012}
{\sc Ding, R., Fu, Q., Lou, J.~G., Lin, Q., Zhang, D., Shen, J., and Xie, T.}
\newblock Healing online service systems via mining historical issue repositories.
\newblock {\em ASE\/} (2012), 318--321.

\bibitem{Ding2014}
{\sc Ding, R., Fu, Q., Lou, J.~G., Lin, Q., Zhang, D., and Xie, T.}
\newblock Mining historical issue repositories to heal large-scale online service systems.
\newblock {\em Proceedings of the Int. Conf. on Dependable Systems and Networks\/} (9 2014), 311--322.

\bibitem{Durieux2017}
{\sc Durieux, T., Hamadi, Y., and Monperrus, M.}
\newblock Production-driven patch generation.
\newblock {\em ICSE\/} (6 2017), 23--26.

\bibitem{Ford2018}
{\sc Ford, S., and Olmsted, A.}
\newblock Security vulnerabilities in javascript hotpatching in ios with a commercial and open-source tool.
\newblock {\em Int. Conf. on Information Society 2018-January\/} (5 2018), 108--110.

\bibitem{Fu2012}
{\sc Fu, Q., Lou, J.~G., Lin, Q.~W., Ding, R., Zhang, D., Ye, Z., and Xie, T.}
\newblock Performance issue diagnosis for online service systems.
\newblock {\em Proceedings of the IEEE Symposium on Reliable Distributed Systems\/} (2012), 273--278.

\bibitem{Gao2009}
{\sc Gao, Q., Zhang, W., Tang, Y., and Qin, F.}
\newblock First-aid: Surviving and preventing memory management bugs during production runs.
\newblock {\em Proceedings of the 4th ACM European Conf. on Computer Systems, EuroSys'09\/} (2009), 159--172.

\bibitem{Glerum2009}
{\sc Glerum, K., Kinshumann, K., Greenberg, S., Aul, G., Orgovan, V., Nichols, G., Grant, D., Loihle, G., and Hunt, G.}
\newblock Debugging in the (very) large: Ten years of implementation and experience.
\newblock {\em SOSP'09 - Proceedings of the 22nd ACM SIGOPS Symposium on Operating Systems Principles\/} (2009), 103--116.

\bibitem{Gomez2015}
{\sc Gomez, M., Martineza, M., Monperrus, M., and Rouvoy, R.}
\newblock When app stores listen to the crowd to fight bugs in the wild.
\newblock {\em ICSE 2\/} (8 2015), 567--570.

\bibitem{surveyLeGoues}
{\sc Goues, C.~L., Pradel, M., and Roychoudhury, A.}
\newblock Automated program repair.
\newblock {\em Communications of the ACM\/} (2019).

\bibitem{Gupta2008}
{\sc Gupta, M., Banerjee, S., Agrawal, M., and Rao, H.~R.}
\newblock Security analysis of internet technology components enabling globally distributed workplacesa framework.
\newblock {\em ACM Transactions on Internet Technology (TOIT) 8\/} (10 2008).

\bibitem{Gomez2017}
{\sc Gómez, M., Adams, B., Maalej, W., Monperrus, M., and Rouvoy, R.}
\newblock App store 2.0: From crowdsourced information to actionable feedback in mobile ecosystems.
\newblock {\em IEEE Software 34\/} (3 2017), 81--89.

\bibitem{Han2023}
{\sc Han, S., Baby, D., and Mendelev, V.}
\newblock Residual adapters for targeted updates in rnn-transducer based speech recognition system.
\newblock {\em 2022 IEEE Spoken Language Technology Workshop, SLT 2022 - Proceedings\/} (2023), 160--166.

\bibitem{Hanna2023}
{\sc Hanna, C., and Petke, J.}
\newblock Hot patching hot fixes: Reflection and perspectives.
\newblock {\em ASE\/} (9 2023).

\bibitem{Hassan2017}
{\sc Hassan, S., Shang, W., and Hassan, A.~E.}
\newblock An empirical study of emergency updates for top android mobile apps.
\newblock {\em Empirical Software Engineering 22\/} (2 2017), 505--546.

\bibitem{He2018}
{\sc He, S., Lin, Q., Lou, J.~G., Zhang, H., Lyu, M.~R., and Zhang, D.}
\newblock Identifying impactful service system problems via log analysis.
\newblock {\em ESEC/FSE 18\/} (10 2018), 60--70.

\bibitem{Herzig2014}
{\sc Herzig, K.}
\newblock Using pre-release test failures to build early post-release defect prediction models.
\newblock {\em Proceedings - Int. Symposium on Software Reliability Engineering, ISSRE\/} (12 2014), 300--311.

\bibitem{Huang2005}
{\sc Huang, H., Tsai, W.~T., and Chen, Y.}
\newblock Autonomous hot patching for web-based applications.
\newblock {\em Proceedings - Int. Computer Software and Applications Conf. 2\/} (2005), 51--56.

\bibitem{Huang2016}
{\sc Huang, Z., Dangelo, M., Miyani, D., and Lie, D.}
\newblock Talos: Neutralizing vulnerabilities with security workarounds for rapid response.
\newblock {\em Proceedings - IEEE Symposium on Security and Privacy\/} (8 2016), 618--635.

\bibitem{IEEEThesaurus}
{\sc IEEE}.
\newblock July 2023 ieee thesaurus version 1.02 created by the institute of electrical and electronics engineers (ieee).

\bibitem{Illes-Seifert}
{\sc Illes-Seifert, T., and Paech, B.}
\newblock Exploring the relationship of a file's history and its fault-proneness: An empirical study.
\newblock {\em Testing: Academic and Industrial Conf. Practice and Research Techniques\/} (2008), 13--22.

\bibitem{Islam2023}
{\sc Islam, C., Prokhorenko, V., and Babar, M.~A.}
\newblock Runtime software patching: Taxonomy, survey and future directions.
\newblock {\em Journal of Systems and Software 200\/} (6 2023), 111652.

\bibitem{Jin2011}
{\sc Jin, G., Song, L., Zhang, W., Lu, S., and Liblit, B.}
\newblock Automated atomicity-violation fixing.
\newblock {\em Conf. on Programming Language Design and Implementation\/} (2011), 389--400.

\bibitem{Karale2016}
{\sc Karale, S.~V., and Kaushal, V.}
\newblock An automation framework for configuration management to reduce manual intervention.
\newblock {\em ACM Int. Conf. Proceeding Series 12-13-August-2016\/} (8 2016).

\bibitem{Khomh2011}
{\sc Khomh, F., Chan, B., Zou, Y., and Hassan, A.~E.}
\newblock An entropy evaluation approach for triaging field crashes: A case study of mozilla firefox.
\newblock {\em Working Conf. on Reverse Engineering\/} (2011), 261--270.

\bibitem{Khomh2012}
{\sc Khomh, F., Dhaliwal, T., Zou, Y., and Adams, B.}
\newblock Do faster releases improve software quality? an empirical case study of mozilla firefox.
\newblock {\em IEEE Int. Working Conf. on Mining Software Repositories\/} (2012), 179--188.

\bibitem{Kolassa2013}
{\sc Kolassa, C., Riehle, D., and Salim, M.~A.}
\newblock The empirical commit frequency distribution of open source projects.
\newblock {\em Proceedings of the Int. Symposium on Open Collaboration\/} (2013).

\bibitem{Li2019}
{\sc Li, F., Chetty, M., Rogers, L., Mathur, A., and Malkin, N.}
\newblock Keepers of the machines: Examining how system administrators manage software updates.
\newblock {\em Fifteenth Symposium on Usable Privacy and Security (SOUPS 2019)\/} (2019), 273--288.

\bibitem{Li2011}
{\sc Li, Z., and Long, J.}
\newblock A case study of measuring degeneration of software architectures from a defect perspective.
\newblock {\em Proceedings - Asia-Pacific Software Engineering Conf., APSEC\/} (2011), 242--249.

\bibitem{Lin2017}
{\sc Lin, D., Bezemer, C.~P., and Hassan, A.~E.}
\newblock Studying the urgent updates of popular games on the steam platform.
\newblock {\em Empirical Software Engineering 22\/} (8 2017), 2095--2126.

\bibitem{Lin2007}
{\sc Lin, Z., Jiang, X., Xu, D., Mao, B., and Xie, L.}
\newblock Autopag: Towards automated software patch generation with source code root cause identification and repair.
\newblock {\em eProceedings of the 2nd ACM Symposium on Information, Computer and Communications Security, ASIACCS '07\/} (2007), 329--340.

\bibitem{Lou2013}
{\sc Lou, J.~G., Lin, Q., Ding, R., Fu, Q., Zhang, D., and Xie, T.}
\newblock Software analytics for incident management of online services: An experience report.
\newblock {\em ASE\/} (2013), 475--485.

\bibitem{Luo2018}
{\sc Luo, L., Nath, S., Sivalingam, R., Musuvathi, M., and Ceze, L.}
\newblock Troubleshooting transiently-recurring errors in production systems with blame-proportional logging troubleshooting transiently-recurring problems in production systems with blame-proportional logging.
\newblock {\em USENIX Annual Technical Conf. (USENIX ATC 18)\/} (2018), 321--334.

\bibitem{MacHiry2020}
{\sc MacHiry, A., Redini, N., Camellini, E., Kruegel, C., and Vigna, G.}
\newblock Spider: Enabling fast patch propagation in related software repositories.
\newblock {\em Proceedings - IEEE Symposium on Security and Privacy 2020-May\/} (5 2020), 1562--1579.

\bibitem{Malone2021}
{\sc Malone, M., Wang, Y., Snow, K., and Monrose, F.}
\newblock Applicable micropatches and where to find them: Finding and applying new security hot fixes to old software.
\newblock {\em Proc. - 2021 IEEE 14th Int. Conf. on Software Testing, Verification and Validation, ICST 2021\/} (4 2021), 394--405.

\bibitem{Marconato2012}
{\sc Marconato, G.~V., Nicomette, V., and Kaâniche, M.}
\newblock Security-related vulnerability life cycle analysis.
\newblock {\em Int. Conf. on Risks and Security of Internet and Systems\/} (2012).

\bibitem{Marra2020}
{\sc Marra, M., Polito, G., and Boix, E.~G.}
\newblock A debugging approach for live big data applications.
\newblock {\em Science of Computer Programming 194\/} (8 2020).

\bibitem{Mockus2000}
{\sc Mockus, A., Fielding, R.~T., Herbsleb, J., Labs, B., and Blvd, S.}
\newblock A case study of open source software development: The apache server.
\newblock {\em ICSE\/} (2000).

\bibitem{Mulliner2013}
{\sc Mulliner, C., Oberheide, J., Robertson, W., and Kirda, E.}
\newblock Patchdroid: Scalable third-party security patches for android devices.
\newblock {\em ACM Int. Conf. Proceeding Series\/} (2013), 259--268.

\bibitem{Novark2007}
{\sc Novark, G., Berger, E.~D., and Zorn, B.~G.}
\newblock Exterminator: Automatically correcting memory errors with high probability.
\newblock {\em Conf. on Programming Language Design and Implementation\/} (2007), 1--11.

\bibitem{Oosterhuis2021}
{\sc Oosterhuis, H., and Rijke, M. D.~D.}
\newblock Robust generalization and safe query-specializationin counterfactual learning to rank.
\newblock {\em The Web Conference 2021 - Proceedings of the World Wide Web Conference, WWW 2021\/} (4 2021), 158--170.

\bibitem{Pamunuwa2023}
{\sc Pamunuwa, V., Deraniyagala, D., Kulasekara, V., Thennakoon, R., and Lankasena, B.}
\newblock Investigating the impact of software maintenance activities on software quality: Case study.

\bibitem{Parameshwaran2015}
{\sc Parameshwaran, I., Budianto, E., Shinde, S., Dang, H., Sadhu, A., and Saxena, P.}
\newblock Auto-patching dom-based xss at scale.
\newblock {\em ESEC/FSE\/} (8 2015), 272--283.

\bibitem{Payer2013}
{\sc Payer, M., and Gross, T.~R.}
\newblock Hot-patching a web server: A case study of asap code repair.
\newblock {\em Annual Conf. on Privacy, Security and Trust\/} (2013), 143--150.

\bibitem{Perkins2009}
{\sc Perkins, J.~H., Kim, S., Larsen, S., Amarasinghe, S., Bachrach, J., Carbin, M., Pacheco, C., Sherwood, F., Sidiroglou, S., Sullivan, G., Wong, W.~F., Zibin, Y., Ernst, M.~D., and Rinard, M.}
\newblock Automatically patching errors in deployed software.
\newblock {\em Proceedings of the ACM SIGOPS Symposium on Operating Systems Principles\/} (2009), 87--102.

\bibitem{Pozo2019}
{\sc Pozo, F., and Rodriguez-Navas, G.}
\newblock A semi-distributed self-healing protocol for run-time repairs of time-triggered schedules.
\newblock {\em IEEE Int. Conf. on Emerging Technologies and Factory Automation, ETFA 2019-September\/} (9 2019), 1399--1402.

\bibitem{Qin2005}
{\sc Qin, F., Tucek, J., Sundaresan, J., and Zhou, Y.}
\newblock Rx: Treating bugs as allergies - a safe method to survive software failures.
\newblock {\em Proceedings of the 20th ACM Symposium on Operating Systems Principles, SOSP 2005\/} (2005), 235--248.

\bibitem{Qin2008}
{\sc Qin, L., Li, Y., and Yue, C.}
\newblock Dataflow analysis for known vulnerability prevention system.
\newblock {\em 2008 IEEE Int. Conf. on Cybernetics and Intelligent Systems, CIS 2008\/} (2008), 1032--1035.

\bibitem{Ramaswamy2010}
{\sc Ramaswamy, A., Bratus, S., Smith, S.~W., and Locasto, M.~E.}
\newblock Katana: A hot patching framework for elf executables.
\newblock {\em ARES 2010 - 5th Int. Conf. on Availability, Reliability, and Security\/} (2010), 507--512.

\bibitem{Rasche2008}
{\sc Rasche, A., and Polze, A.}
\newblock Redac - dynamic reconfiguration of distributed component-based applications with cyclic dependencies.
\newblock {\em Proceedings - IEEE Symposium on Object/Component/Service-Oriented Real-Time Distributed Computing\/} (2008), 322--330.

\bibitem{Russinovich2021}
{\sc Russinovich, M., Govindaraju, N., Raghuraman, M., Hepkin, D., Schwartz, J., and Kishan, A.}
\newblock Virtual machine preserving host updates for zero day patching in public cloud.
\newblock {\em Proceedings of the European Conf. on Computer Systems 21\/} (4 2021), 114--129.

\bibitem{Saieva2020}
{\sc Saieva, A., and Kaiser, G.}
\newblock Binary quilting to generate patched executables without compilation.
\newblock {\em ACM Workshop on Forming an Ecosystem Around Software Transformation\/} (11 2020), 3--8.

\bibitem{Saieva2022}
{\sc Saieva, A., and Kaiser, G.}
\newblock Update with care: Testing candidate bug fixes and integrating selective updates through binary rewriting.
\newblock {\em Journal of Systems and Software 191\/} (9 2022), 111381.

\bibitem{Salehi2022}
{\sc Salehi, M., and Pattabiraman, K.}
\newblock Poster autopatch: Automatic hotpatching of real-time embedded devices.
\newblock {\em Proceedings of the ACM Conf. on Computer and Communications Security\/} (11 2022), 3451--3453.

\bibitem{Sarabi2017}
{\sc Sarabi, A., Zhu, Z., Xiao, C., Liu, M., and Dumitraş, T.}
\newblock Patch me if you can: A study on the effects of individual user behavior on the end-host vulnerability state.
\newblock {\em Lecture Notes in Computer Science (including subseries Lecture Notes in Artificial Intelligence and Lecture Notes in Bioinformatics) 10176 LNCS\/} (2017), 113--125.

\bibitem{Savor2016}
{\sc Savor, T., Douglas, M., Gentili, M., Williams, L., Beck, K., and Stumm, M.}
\newblock Continuous deployment at {Facebook and OANDA}.
\newblock {\em ICSE\/} (5 2016), 21--30.

\bibitem{Shen2017}
{\sc Shen, S., Lu, X., Hu, Z., and Liu, X.}
\newblock Towards release strategy optimization for apps in google play.
\newblock {\em ACM Int. Conf. Proceeding Series Part F130951\/} (9 2017).

\bibitem{Shen2020}
{\sc Shen, Z., and Chen, S.}
\newblock A survey of automatic software vulnerability detection, program repair, and defect prediction techniques.
\newblock {\em Security and Communication Networks\/} (2020).

\bibitem{Shihab2011}
{\sc Shihab, E., Mockus, A., Kamei, Y., Adams, B., and Hassan, A.~E.}
\newblock High-impact defects: A study of breakage and surprise defects.
\newblock {\em SIGSOFT/FSE\/} (2011), 300--310.

\bibitem{Sidiroglou2007}
{\sc Sidiroglou, S., Ioannidis, S., and Keromytis, A.~D.}
\newblock Band-aid patching.
\newblock {\em Workshop on Hot Topics in System Dependability\/} (2007), 102--106.

\bibitem{Sidiroglou2005}
{\sc Sidiroglou, S., Locasto, M.~E., Boyd, S.~W., and Keromytis, A.~D.}
\newblock Building a reactive immune system for software services.
\newblock {\em Proceedings of the 2005 USENIX Annual Technical Conf.\/} (2005), 149--161.

\bibitem{Sun2018}
{\sc Sun, D., Fekete, A., Gramoli, V., Li, G., Xu, X., and Zhu, L.}
\newblock R2c: Robust rolling-upgrade in clouds.
\newblock {\em IEEE Transactions on Dependable and Secure Computing 15\/} (9 2018), 811--823.

\bibitem{Tang2012}
{\sc Tang, J., Kim, H., Mascolo, C., and Musolesi, M.}
\newblock Stop: Socio-temporal opportunistic patching of short range mobile malware.
\newblock {\em 2012 IEEE Int. Symposium on a World of Wireless, Mobile and Multimedia Networks, WoWMoM 2012 - Digital Proceedings\/} (2012).

\bibitem{Thota2020}
{\sc Thota, M.~K., Shajin, F.~H., and Rajesh, P.}
\newblock Survey on software defect prediction techniques.
\newblock {\em Int. Journal of Applied Science and Engineering 17\/} (2020), 331--344.

\bibitem{Trivedi2011}
{\sc Trivedi, K.~S., Mansharamani, R., Kim, D.~S., Grottke, M., and Nambiar, M.}
\newblock Recovery from failures due to mandelbugs in it systems.
\newblock {\em Proceedings of IEEE Pacific Rim Int. Symposium on Dependable Computing, PRDC\/} (2011), 224--233.

\bibitem{Truelove2021}
{\sc Truelove, A., de~Almeida, E.~S., and Ahmed, I.}
\newblock We'll fix it in post: What do bug fixes in video game update notes tell us?
\newblock {\em ICSE\/} (5 2021), 736--747.

\bibitem{triage}
{\sc Tucek, J., Lu, S., Huang, C., Xanthos, S., and Zhou, Y.}
\newblock Triage: Diagnosing production run failures at the user's site.
\newblock {\em Operating Systems Review (ACM)\/} (2007), 131--144.

\bibitem{Tucek2007}
{\sc Tucek, J., Lu, S., Huang, C., Xanthos, S., Zhou, Y., Newsome, J., Brumley, D., and Song, D.}
\newblock Sweeper: A lightweight end-to-end system for defending against fast worms.
\newblock {\em Operating Systems Review (ACM)\/} (2007), 115--128.

\bibitem{TowardsJIT}
{\sc Tunde-Onadele, O., Carolina, N., Lin, Y., He, J., and Gu, X.}
\newblock Toward just-in-time patching for containerized applications.
\newblock {\em Proceedings of the 7th Symposium on Hot Topics in the Science of Security\/} (2020).

\bibitem{Olufogorehan2020}
{\sc Tunde-Onadele, O., Lin, Y., He, J., and Gu, X.}
\newblock Self-patch: Beyond patch tuesday for containerized applications.
\newblock {\em Int. Conf. on Autonomic Computing and Self-Organizing Systems\/} (8 2020), 21--27.

\bibitem{Van2005}
{\sc {Van Der Storm}, T.}
\newblock Continuous release and upgrade of component-based software.
\newblock {\em Proceedings of the 12th Int. Workshop on Software Configuration Management, SCM 2005\/} (2005), 43--57.

\bibitem{storm2007}
{\sc {Van Der Storm}, T.}
\newblock Binary change set composition.
\newblock {\em Lecture Notes in Computer Science (including subseries Lecture Notes in Artificial Intelligence and Lecture Notes in Bioinformatics) 4608 LNCS\/} (2007), 17--32.

\bibitem{Vaniea2016}
{\sc Vaniea, K., and Rashidi, Y.}
\newblock Tales of software updates: The process of updating software.
\newblock {\em Conf. on Human Factors in Computing Systems - Proceedings\/} (5 2016), 3215--3226.

\bibitem{Verma2017}
{\sc Verma, S., and Roy, S.}
\newblock Synergistic debug-repair of heap manipulations.
\newblock {\em FSE Part F130154\/} (8 2017), 163--173.

\bibitem{Vogler2016}
{\sc Vögler, M., Schleicher, J.~M., Inzinger, C., and Dustdar, S.}
\newblock A scalable framework for provisioning large-scale iot deployments.
\newblock {\em ACM Transactions on Internet Technology 16\/} (3 2016).

\bibitem{Vogler2015}
{\sc Vögler, M., Schleicher, J.~M., Inzinger, C., Nastic, S., Sehic, S., and Dustdar, S.}
\newblock Leonore - large-scale provisioning of resource-constrained iot deployments.
\newblock {\em Proceedings - IEEE Int. Symposium on Service-Oriented System Engineering 30\/} (6 2015), 78--87.

\bibitem{Wang2022}
{\sc Wang, Y., Jiang, S., and Cui, B.}
\newblock Tjosconf: Automatic and safe system environment operations platform.
\newblock {\em ACM Int. Conf. Proceeding Series 2022\/} (2 2022), 21--28.

\bibitem{tightRelease}
{\sc Weiß, C., Premraj, R., Zimmermann, T., and Zeller, A.}
\newblock How long will it take to fix this bug?
\newblock {\em Proceedings - ICSE 2007 Workshops: Fourth International Workshop on Mining Software Repositories, MSR 2007\/} (2007).

\bibitem{Woodard2012}
{\sc Woodard, D.~B., and Goldszmidt, M.}
\newblock Online model-based clustering for crisis identification in distributed computing.
\newblock {\em Journal of the American Statistical Association 106\/} (3 2012), 49--60.

\bibitem{Xu2020Source}
{\sc Xu, Z.}
\newblock Source code and binary level vulnerability detection and hot patching.
\newblock {\em ASE\/} (2 2020), 1397--1399.

\bibitem{Xu2020Automatic}
{\sc Xu, Z., Zhang, Y., Zheng, L., Xia, L., Bao, C., X-Lab, B., Wang, Z., Liu, Y., Longri, B. X.-L., Baidu, Z., Liangzhao, X.-L., Baidu, X., Chenfu, X.-L., Baidu, B., and Wang, X.-L.~Z.}
\newblock Automatic hot patch generation for android kernels.
\newblock {\em Proceedings of the USENIX Conf. on Security Symposium\/} (2020).

\bibitem{Yuan2006}
{\sc Yuan, C., Ma, W.~Y., Wen, J.~R., Li, J., Zhang, Z., and Wang, Y.~M.}
\newblock Automated known problem diagnosis with event traces.
\newblock {\em ACM SIGOPS Operating Systems Review 40\/} (4 2006), 375--388.

\bibitem{Zhang2013}
{\sc Zhang, H., Gong, L., and Versteeg, S.}
\newblock Predicting bug-fixing time: An empirical study of commercial software projects.
\newblock {\em ICSE\/} (2013), 1042--1051.

\bibitem{Zhang2018}
{\sc Zhang, H., and Qian, Z.}
\newblock Precise and accurate patch presence test for binaries.
\newblock {\em 27th USENIX Security Symposium\/} (2018).

\bibitem{Zhang2015}
{\sc Zhang, H., Zhao, L., Xu, L., Wang, L., and Wu, D.}
\newblock Vpatcher: Vmi-based transparent data patching to secure software in the cloud.
\newblock {\em TrustCom\/} (1 2015), 943--948.

\bibitem{Zhang2014}
{\sc Zhang, M., and Yin, H.}
\newblock Appsealer: Automatic generation of vulnerability-specific patches for preventing component hijacking attacks in android applications.
\newblock {\em NDSS 14\/} (2014), 23--26.

\bibitem{Zhang2005}
{\sc Zhang, S., Cohen, I., Goldszmidt, M., Symons, J., and Fox, A.}
\newblock Ensembles of models for automated diagnosis of system performance problems.
\newblock {\em Proceedings of the Int. Conf. on Dependable Systems and Networks\/} (2005), 644--653.

\bibitem{Zhang2021}
{\sc Zhang, X., Xu, Y., Qin, S., He, S., Qiao, B., Li, Z., Zhang, H., Li, X., Dang, Y., Lin, Q., Chintalapati, M., Rajmohan, S., and Zhang, D.}
\newblock Onion: Identifying incident-indicating logs for cloud systems.
\newblock {\em ESEC/FSE 21\/} (8 2021), 1253--1263.

\bibitem{Zhang2017}
{\sc Zhang, X., Zhang, Y., Li, J., Hu, Y., Li, H., and Gu, D.}
\newblock Embroidery: Patching vulnerable binary code of fragmentized android devices.
\newblock {\em Int. Conf. on Software Maintenance and Evolution\/} (11 2017), 47--57.

\bibitem{Zhou2015}
{\sc Zhou, H., Lou, J.-G., Zhang, H., Lin, H., Lin, H., and Qin, T.}
\newblock An empirical study on quality issues of production big data platform.
\newblock {\em ICSE\/} (2015).

\bibitem{Zhou2020}
{\sc Zhou, L., Zhang, F., Liao, J., Ning, Z., Xiao, J., Leach, K., Weimer, W., and Wang, G.}
\newblock Kshot: Live kernel patching with smm and sgx.
\newblock {\em Proceedings - 50th Annual IEEE/IFIP International Conference on Dependable Systems and Networks, DSN 2020\/} (6 2020), 1--13.

\end{thebibliography}

\end{document}